\documentclass[12pt]{article}

\usepackage[top=1in, bottom=1in, left=1in, right=1in]{geometry}

\usepackage{pdflscape}

\usepackage{hyperref}
\usepackage{titlesec}
\titlespacing\section{0pt}{12pt}{0pt}
\titlespacing\subsection{0pt}{6pt}{0pt}
\titlespacing\subsubsection{0pt}{6pt}{0pt}

\usepackage{titletoc}
\usepackage{lipsum}

\usepackage{lineno} 
\usepackage{times}  

\usepackage[table]{xcolor}
\usepackage[normalem]{ulem}
\usepackage{longtable,booktabs} 
\usepackage{footnote}
\usepackage{tablefootnote}  
\usepackage{capt-of}
\usepackage{xcolor}
\usepackage{booktabs}
\usepackage{amsmath}
\usepackage{amsthm}
\usepackage{amssymb}
\usepackage{bbold} 

\usepackage{multicol,multirow,makecell}

\usepackage{url}
\usepackage{threeparttable}
\usepackage{lscape}
\usepackage{longtable,booktabs} 
\usepackage{geometry}
\usepackage{verbatim}
\usepackage{enumitem}
\usepackage{titlesec}
\usepackage{tabularx}
\usepackage{tikz}
\usepackage{subcaption}
\usepackage{graphicx}
\usepackage{float}
\usepackage[bottom]{footmisc}

\usepackage{hhline}

\usepackage[numbers]{natbib}
 \usepackage{ulem}
 \usepackage{soul}
 \usepackage{tabularray}
\UseTblrLibrary{booktabs}
\usepackage{hyperref}

\usepackage{setspace}
\usepackage{appendix}
\usepackage{xcolor}

\titlespacing\section{0pt}{10pt}{10pt}  
\titlespacing\subsection{0pt}{8pt}{8pt}  

\hypersetup{
	colorlinks   = true,  
	urlcolor     = blue,  
	linkcolor    = blue,  
	citecolor   = blue     
}

\usepackage{hyperref}

\usepackage{array}
\newcommand{\PreserveBackslash}[1]{\let\temp=\\#1\let\\=\temp}
\newcolumntype{C}[1]{>{\PreserveBackslash\centering}p{#1}}
\newcolumntype{R}[1]{>{\PreserveBackslash\raggedleft}p{#1}}
\newcolumntype{L}[1]{>{\PreserveBackslash\raggedright}p{#1}}

\definecolor{mypink1}{RGB}{255, 204, 255}
\definecolor{mypurple1}{RGB}{204,51,255}
\definecolor{mygreen1}{RGB}{61,245,0}
\definecolor{myorange1}{RGB}{255,102,0}
\definecolor{mymelon}{RGB}{254,136,99}
\newcommand*{\rom}[1]{\expandafter\@slowromancap\romannumeral #1@}

\newcommand{\hide}[1]{} %hide

\newcommand{\note}[1]{[\ul{\textbf{#1}}]}

\definecolor{mygreen1}{RGB}{56, 183, 0}
\definecolor{myblue1}{RGB}{44, 131, 195}
\definecolor{myred1}{RGB}{214, 39, 40}
\definecolor{mypurple1}{RGB}{	154, 0, 145}

\theoremstyle{plain}

\theoremstyle{definition}

\newcommand{\RNum}[1]{\uppercase\expandafter{\romannumeral #1\relax}}

\begin{document}

\title{When Experimental Economics Meets \\Large Language Models: Evidence-based Tactics\thanks{Wang, Yao and Zhang are the co-first authors on the paper (alphabetically ordered). Liu and Zhong are corresponding authors. Wang, Gai and Liu: Department of Economics, School of Economics and Management, Tsinghua University (Wang: shu-wang20@mails.tsinghua.edu.cn; Gai: gaijn23@mails.tsinghua.edu.cn; Liu: liuxiao@sem.tsinghua.edu.cn). Yao: Department of Computer Science and Technology, Tsinghua University (yaozj20@mails.tsinghua.edu.cn). Zhang: PBC School of Finance, Tsinghua University (zhangshh.20@pbcsf.tsinghua.edu.cn). Zhong: Department of Economics, National University of Singapore (zhongsongfa@gmail.com ). Liu gratefully acknowledges financial support by NSFC (72222005, 72342032). We thank Chuhan Zhang for her excellent research assistance.}
}
\author{Shu Wang \and Zijun Yao \and Shuhuai Zhang \and Jianuo Gai \and Tracy Xiao Liu \and Songfa Zhong}

\date{\today}
\maketitle
\begin{abstract}

Advancements in large language models (LLMs) have sparked a growing interest in measuring and understanding their behavior through experimental economics. However, there is still a lack of established guidelines for designing economic experiments for LLMs. Inspired by principles from experimental economics with insights from LLM research in artificial intelligence, we outline key considerations in the experimental design and implementation stage, and perform two sets of experiments to assess the impact of these considerations on LLMs' responses. Based on our findings, we discuss seven practical tactics for conducting experiments with LLMs. Our study enhances the design, replicability, and generalizability of LLM experiments, and broadens the scope of experimental economics in the digital age.

\end{abstract}

\medskip

\textit{Keywords}: large language models (LLMs), experimental economics, decision-making, \\ rationality, preference

%********************************
\newpage

\section{Introduction}
The rise of large language models (LLMs), particularly since the launch of ChatGPT in November 2022, has sparked significant interest in understanding their capacities \citep{chui2023economic,Reuters2023}. Research has shown that LLMs exhibit impressive abilities beyond mere language generation, including generating computer code, solving complex math problems, and demonstrating human-like reasoning. This growing prominence has led to a surge in studies that examine how LLMs perform in economic experiments using decision-making tasks---referred to as ``LLM experiments". These studies measure the preferences, rationality, heuristics, and biases of LLMs, and assess whether they can pass a form of the Turing test that evaluates behavioral similarity to humans \citep{chen2023emergence,horton2023large,leng2023llm,mei2024turing}. As LLMs continue to be integrated into various aspects of daily life and reshape our interactions with technology, these studies are becoming increasingly crucial and popular \citep{meng2024ai}. However, despite this rapid progress, little has been done to examine the underlying rules of conducting economic experiments in the context of LLMs.

This study investigates the rules of LLM experiments. Evaluating LLM experiments requires a different perspective compared to traditional natural language processing (NLP) benchmarks. In typical NLP benchmarks—such as language translation, mathematical reasoning, and code generation—there exist correct answers \citep{cobbe2021gsm8k, chen2021evaluating, papineni2002bleu}. 
The objective of prompt engineering in these scenarios is to improve the accuracy of LLMs' outputs. In contrast, the main objective of LLM experiments is to measure the behavior of LLMs and to assess whether they exhibit human-like behavior \citep{leng2023llm,goli2024frontiers,leng2024can,leng2024reduce,liu2024large,deng2025can,fedyk2025chatgpt,sun2025large,zhu2025language}, a domain where the outputs do not have definitive right or wrong answers. Consequently, while LLMs demonstrate robustness to minor prompt variations in standard benchmark tasks, the influence of prompt formulation on complex decision-making tasks remains largely unexplored \citep{wei2022chain,razavi2025benchmarking}. This issue may affect the robustness and generalizability of the conclusions drawn from such experiments.

To address this issue, we synthesize principles from experimental economics with insights derived from LLM research in artificial intelligence. 
First, it is natural to evaluate the results of LLM experiments using established principles in experimental economics. For example, experimental economists emphasize the clarity and neutrality of instructions to ensure that experiments align closely with theoretical models, as well as the importance of monetary incentives and the incentive compatibility of mechanisms \citep{smith1982microeconomic,davis1993experimental,friedman1994experimental}. Second, the unique characteristics of LLM experiments introduce new questions for experimental economics. For example, when addressing multiple decisions, researchers can either consolidate them into a single prompt (single-turn dialogue) or separate them into distinct prompts while incorporating prior interactions as memory (multi-turn dialogue). The lack of standardization regarding the choice between single-turn and multi-turn dialogues in existing LLM experiments further complicates the evaluation process. Therefore, it is essential to systematically identify and consider the characteristics, parameters, and procedures at each stage of LLM experiments that can influence the results. Although universal rules may not be applicable, some common practices can provide valuable guidelines for future research. 

In this paper, we design two sets of experiments to examine the effect of some key considerations in experimental protocols on LLMs' responses from different perspectives.
In Case Study~1, we examine the economic rationality of LLMs \citep{chen2023emergence,kim2024learning}, in which LLMs make a series of budgetary decisions \citep{andreoni2002giving,choi2007consistency,chen2023emergence}. In Case Study~2, we examine the economic preferences of LLMs \citep{mei2024turing,xie2024different}, in which LLMs make decisions in a series of behavioral games, including the dictator game, ultimatum game, public goods game and bomb risk game. We conduct the experiments using four popular LLMs, including GPT-4o-2024-11-20 (hereafter referred to as GPT), DeepSeek-V3-0324 (DeepSeek), Llama-3.1-8B-Instruct (Llama) and Qwen2.5-7B-Instruct (Qwen).
We find that assigned personas, such as occupation, significantly affect preferences but not rationality. Second, the single-turn dialogue approach significantly decreases the rationality of Llama and Qwen, while that of GPT and DeepSeek is not affected by the dialogue type. Third, limiting responses to multiple-choice options instead of allowing open-ended responses reduces the rationality of Llama and Qwen. It also significantly changes the outputs of all four LLMs in approximately half of the behavioral games. 
The results of two case studies inform actionable tactics, and shed light on the significance of design, replicability, and generalizability in LLM experiments.  
In a broad sense, our study contributes to the replicability of experimental studies by providing a framework that can be applied consistently and be comparable across experiments, thereby strengthening the reliability and validity of experimental findings \citep{camerer2016evaluating,camerer2018evaluating}.

\section{Experimental Economics: from Humans to LLMs}
\label{sec:principle}

In this section, we begin by outlining foundational principles drawn from the practices of human experiments within experimental economics, then review the procedures of existing research on LLM experiments. In particular, we summarize several key principles for conducting human experiments by reviewing classical textbooks and survey papers \citep{davis1993experimental,friedman1994experimental,croson2005method}. In the experimental design stage, these include (1) follow the induced value theory, (2) avoid deception, and (3) minimize experimenter demand effects. In the experimental implementation stage, the principles are (1) use context-neutral instructions, (2) make instructions comprehensible, (3) provide illustrative examples, (4) incorporate comprehension assessments, and (5) prevent inter-subject communication. We detail these principles in Supplementary Section~\ref{sec:human exp}.  Moreover, we also provide a comprehensive overview of the development of LLMs, introduce various types of LLMs, and outline the procedures to prompt them in Supplementary Section~\ref{app:overview}.

Next, we summarize 12 papers involving LLM experiments that have been published in leading journals or conferences (Supplementary Table~\ref{tab:llms_exp_sum}), including \textit{Proceedings of the National Academy of Sciences} (\textit{PNAS}), \textit{Nature Human Behaviour}, \textit{Marketing Science}, and \textit{Manufacturing \& Service Operations Management}, among others. Although there is a rapidly growing literature on this topic \citep[e.g., ][]{leng2023llm,kim2024learning,xie2024different,deng2025can}, we decide to include only published papers because their experimental procedures are finalized and many of them provide the code which enables us to replicate their results. We apply what we have learned from both artificial intelligence and human experiments to evaluate these studies. We find that these studies exhibit a large degree of heterogeneity in the choice of parameters in the prompts (Supplementary Table~\ref{tab:llms_exp_eva}), as summarized below. 

\begin{enumerate}[itemsep=3pt,parsep=0pt,topsep=3pt,partopsep=3pt]
\item \textit{Temperature.} 
Temperature is a unique design parameter raised by LLM experiments. Take the models in GPT family as an example, three out of twelve studies set temperature at 0, five choose 1, one chooses 0.7, and two studies have tried multiple values such as 0 and 1 and do not find significant differences across temperatures \citep{chen2023emergence, kosinski2024evaluating}.

\item \textit{Persona.} 
Prior experimental studies have shown that individual choices are affected by their demographics \citep{croson2009gender,dohmen2011individual}. Although most LLM experiments evaluate the performance of LLMs without assigning specific demographics, some examine the impact of demographics with different purposes. To compare with the choices of undergraduate students in the games, Brookins and DeBacker  \citep{brookins2024playing} explicitly assign LLMs the persona of an undergraduate student. To explore whether LLMs can exhibit human-like behavioral heterogeneity across groups, Chen et al. \citep{chen2023emergence} and Mei et al. \citep{mei2024turing} vary different demographics such as gender, ethnicity, and occupation. 

\item \textit{Incentive.} Four studies incorporate a hypothetical incentive scheme into the prompt, while six do not. The remaining two studies apply the incentive to some tasks and omit it in others.

\item \textit{Illustrative example and understanding question.} 
Nearly all LLM experiments do not include examples and understanding questions. We conjecture this is because the tasks used in these studies are simple and easy to understand. 
However, we notice the use of illustrative examples \citep{mei2024turing,wangwill2024} and understanding questions \citep{chen2023emergence} in some studies. Take the latter as an example, the rationale lies in its goal to compare the performance of LLMs with human subjects, where understanding questions are employed in human experiments.

\item \textit{Single-turn vs. multi-turn dialogue.} 
Among these studies, four of them involve situations that each LLM agent needs to answer multiple questions \citep{binz2023using,chen2023emergence,webb2023emergent,mei2024turing},
while the remaining studies ask each agent only to answer one question. In practice, both single-turn dialogue and multi-turn dialogue can be applied to instruct LLMs to collect answers for multiple questions. Supplementary Figure~\ref{fig:number of turns} provides an illustration of these methods. Within these three studies, 
one utilizes the single-turn dialogue, while the three two employ the multi-turn dialogue.
Notably, Mei et al. \citep{mei2024turing} pose multiple questions to each LLM agent and provide feedback to elicit its strategy in several games. Such feedback can only be delivered using the multi-turn dialogue, thus single-turn dialogue is not feasible due to the experimental design.

\item \textit{Open-ended vs. multiple-choice answer type.} 
Open-ended answer type prompts responses without predefined options, whereas multiple-choice type explicitly lists a finite set of choices for LLMs to select from within the prompts. Of the studies reviewed, two use the open-ended answer type, one provides multiple choices, and the remaining nine incorporate both types.

\item \textit{Invalid answer.}
There are instances in which models fail to generate valid answers. For example, LLMs may produce statements such as ``As an AI language model, I cannot participate in surveys nor accept tokens as payment.” \citep{goli2024frontiers} or ``As an AI language model, I am not capable of making decisions on my own.” \citep{chen2023emergence}. Another type of invalid answers features that LLMs generate elaborative explanatory content without explicitly providing the requested outputs. In handling these invalid responses, four studies report the proportion of them in the analysis, whereas the others do not mention the proportion.
\end{enumerate}

\section{Case Study 1: LLMs' Economic Rationality}
\label{sec_4_rationality}
In this section, we implement the budgetary tasks of Chen et al. \citep{chen2023emergence} to examine the extent to which the considerations discussed above influence the performance of LLMs.

\subsection{Experimental Design} 
We choose four representative LLMs to carry out our tasks: 
(1) GPT-4o-2024-11-20, the advanced LLM developed by OpenAI, which is widely regarded as the pioneering leader in the field, 
(2) DeepSeek-V3-0324, the open-sourced model with the largest parameter size \citep{aiindex2025},
(3) Llama-3.1-8B-Instruct, a widely studied model in artificial intelligence and the most downloaded text-generation model released since 2024, ranked by Hugging Face, and 
(4) Qwen2.5-7B-Instruct, an emerging model that has gained increasing attention in the field of artificial intelligence since early 2025 \citep[e.g.,][]{gandhi2025cognitive,muennighoff2025s1,xie2025logic}. All four models are prompted via the application programming interface (API) using Python.
 
\subsubsection{Budgetary Decision Tasks}

Economic rationality is a central assumption in economics and is widely evaluated through various domains of decision-making \citep[e.g.,][]{andreoni2002giving,choi2007consistency,choi2014more,echenique2011money,kim2018role}.
We use standard budgetary decision tasks in risk preference \citep{choi2007consistency} and social preference \citep{andreoni2002giving} to measure economic rationality.
In the risk preference domain, the agent allocates 100 points to two accounts, where the exchange rates between points and payoffs differ. She receives the reward from a random account with probability 50\%. 
For the social preference, the agent allocates 100 points between herself and another randomly matched agent. These points are converted into payoffs with distinct exchange rates for the two agents. We construct 25 rounds of tasks in each simulation, with budget lines that vary between rounds.

\subsubsection{Baseline Condition}
In the baseline condition, we closely align the experimental procedures with those commonly used in human experiments. The temperature is set to the default value of each LLM,\footnote{The default temperatures are 1 for GPT (\href{https://platform.openai.com/docs/api-reference/responses/create}{https://platform.openai.com/docs/api-reference/responses/create}), 0.3 for DeepSeek (\href{https://huggingface.co/deepseek-ai/DeepSeek-V3-0324}{https://huggingface.co/deepseek-ai/DeepSeek-V3-0324}), 0.6 for Llama (\href{https://huggingface.co/meta-llama/Llama-3.1-8B-Instruct/blob/main/generation_config.json}{https://huggingface.co/meta-llama/Llama-3.1-8B-Instruct/blob/main/generation\_config.json}) and 0.7 for Qwen (\href{https://huggingface.co/Qwen/Qwen2.5-7B-Instruct/blob/main/generation_config.json}{https://huggingface.co/Qwen/Qwen2.5-7B-Instruct/blob/main/generation\_config.json}), all last accessed on May 25, 2025.} and there is no specific persona included in the system message. We outline the payment rule in which one round is randomly selected for payment to hypothetically incentivize LLMs. In the user message, we also provide a demonstration example and evaluate the degree of comprehension using understanding questions. 
In each simulation, we use a multi-turn approach, where each round of question is asked separately, with the preceding dialogue included. Thus, each LLM makes one decision per round. Additionally, we use the open-ended answer type and restrict the output to the JSON format to streamline data collection. 
When invalid answers occur, we exclude them and continue until 25 valid answers are collected in a simulation.\footnote{Invalid answers include instances where (1) LLMs refuse to provide answers, (2) the outputs do not adhere to the specified formats, i.e., JSON format in Case Study~1 or number highlighted using ``[[]]'' in Case Study~2, and (3) the answers fail to meet task requirements, e.g., the total points allocated to two accounts do not sum to 100 in Case Study 1 or the number of points falls outside the specified range in both case studies.} Meanwhile, the proportion of invalid answers is also documented.
We repeat the above procedure 100 times to generate 100 simulations, ensuring sufficient statistical power for our subsequent comparisons between conditions.\footnote{By assuming equal variance and specifying the minimum detectable difference in CCEI scores of 0.01, we use the standard deviation obtained in Chen et al. \citep{chen2023emergence} and compute the minimum sample size for each condition, which is 24 in the risk preference and 18 in the social preference, respectively.}

\subsubsection{Experimental Conditions}

Based on the baseline condition, we implement several variations to investigate the impact of the considerations we have summarized. In particular, we focus on the following three factors which suggest that the corresponding variation may influence LLMs' performance. Each condition includes 100 simulations for each model. Supplementary Section~\ref{prompt:case1} provides detailed prompts for the baseline condition and these variations.

\begin{enumerate}[itemsep=3pt,parsep=0pt,topsep=3pt,partopsep=3pt]

    \item \textit{Persona.} We separately assign LLMs a range of demographics, including gender, age, education level and ethnicity as in Chen et al. \citep{chen2023emergence}, as well as occupations as in Mei et al. \citep{mei2024turing}.\footnote{We vary the persona by modifying the role specified in the system prompt, using statements such as ``I want you to act as a male decision maker'' and ``I want you to act as a mathematician'' \citep{chen2023emergence}. For variations in occupations, according to the method of Mei et al. \citep{mei2024turing}, we additionally provide the description of a list of core and supplementary tasks associated with each occupation.} Each specific persona corresponds to a unique condition.
    \item \textit{Dialogue type.} Instead of using multi-turn dialogue, we consider a single-turn dialogue format that incorporates all 25 tasks in one prompt, instructing LLMs to make 25 decisions simultaneously.
    \item \textit{Answer type.} We change the answer type from open-ended to multiple-choice \citep{kim2018role,chen2025how}. Specifically, the continuous budget set is discretized into 21 allocation options with a step size of 5 points, and LLMs are required to select one option for each task.

\end{enumerate}

\subsection{Results for LLMs' Rationality}\label{res:rationality}

In this section, we first report on the economic rationality of LLMs in the baseline condition, followed by the results under experimental conditions. To evaluate the degree of economic rationality, we use the classic measure in revealed preference analysis, the Afriat’s \citep{afriat1972efficiency} critical cost efficiency index (CCEI). CCEI scores range from 0 to 1, with higher scores indicating greater economic rationality. A CCEI score of 1 indicates that the dataset satisfies the generalized axiom of revealed preference (GARP) and is consistent with utility maximization. 

\subsubsection{Rationality in Baseline Condition}

Before discussing the economic rationality of LLMs, we compute the proportion of invalid answers. Among the four models, GPT, DeepSeek and Qwen consistently produce valid answers across all 100 simulations in each domain. Only Llama generates invalid answers in one simulation in the risk preference. These findings suggest that LLMs are generally capable of following instructions and providing the required outputs in budgetary decision tasks.

\begin{figure}[htbp]

\begin{subfigure}{1\textwidth}
    \centering
    \includegraphics[width=0.9\linewidth]{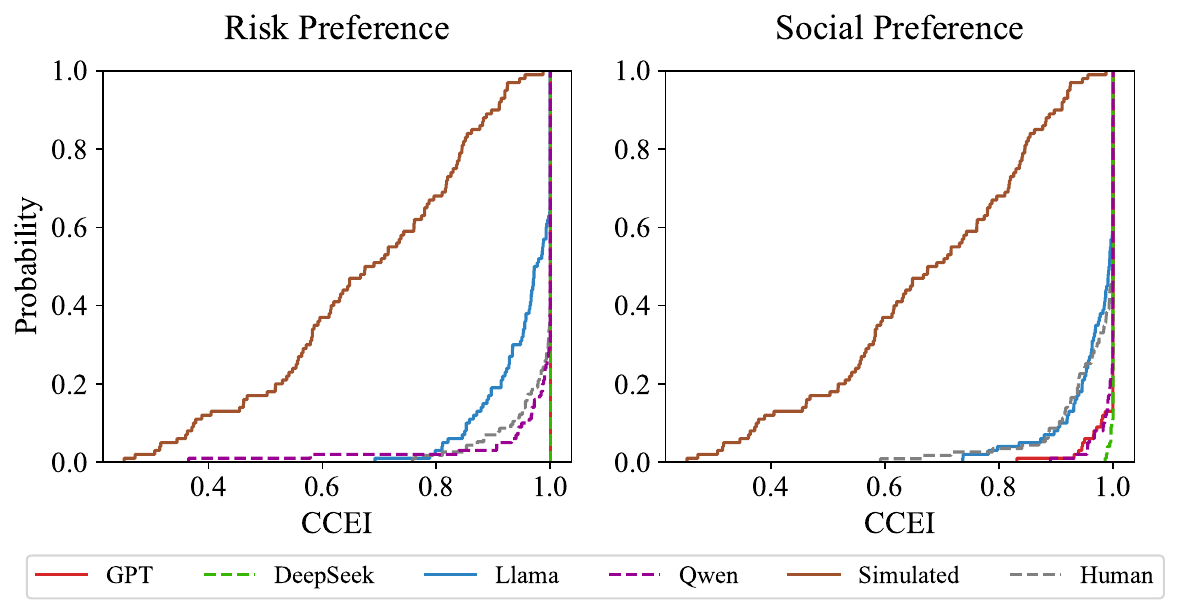}
    \caption{Cumulative Distribution of CCEI Scores}
    \label{fig:rationality—base}
\end{subfigure}

\begin{subfigure}{1\textwidth}
    \centering
    \includegraphics[width=1\linewidth]{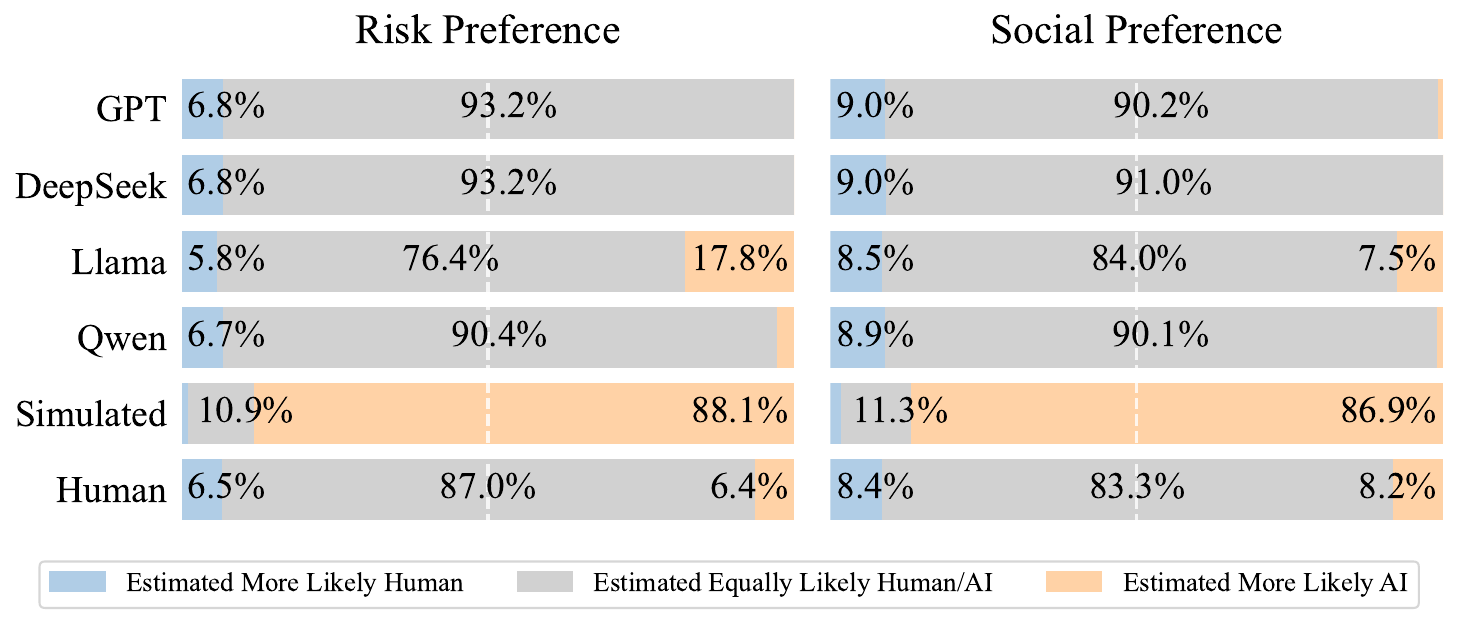}
    \caption{Turing test of CCEI Scores}
    \label{fig:rationality—turing}
\end{subfigure}

\caption{Baseline results of CCEI scores. Human data are from Chen et al. \citep{chen2023emergence}. In Panel b, probabilities below 5\% are not labeled with numbers. The white dashed lines represent the probability of 50\%. In the simulated condition, each comparison involves one CCEI score randomly drawn from the simulated distribution and one from the human distribution. In the human condition, both CCEI scores in each comparison are randomly drawn from the human distribution.}
    \label{fig:case1_base}
\end{figure}

Next, we focus on the economic rationality of LLMs. The average CCEI scores of GPT, DeepSeek, Llama, and Qwen are 1.000, 1.000, 0.953, and 0.980 in the risk preference domain, and 0.994, 0.999, 0.968, and 0.994 in the social preference domain.\footnote{To validate the sufficient power of detecting GARP violations, simulated CCEI is generated using random allocation across 25 decisions for 100 simulated agents \citep{bronars1987power}. The simulated CCEI is significantly lower than all four LLMs in both preference domains (all $p<0.01$).}  
In human experiments using comparable budgetary tasks, the average CCEI ranges from 0.881 to 0.980 in the risk preference \citep{choi2007consistency,choi2014more,halevy2018parametric,chen2023emergence,chen2025how}, and from 0.860 to 0.967 in the social preference \citep{fisman2015distributional,fisman2023distributional,li2017social,li2022experimental,chen2023emergence}. Among the models evaluated, Llama and Qwen may fall within these ranges, while the other two models exhibit higher levels of rationality. We take human decisions in the baseline condition of Chen et al. \citep{chen2023emergence} as a reference and depict the cumulative distribution of CCEI scores for human subjects and the four LLMs in Figure~\ref{fig:rationality—base}. The results indicate that GPT and DeepSeek outperform human subjects in both domains, Qwen surpasses humans in the social preference, while Llama performs worse than humans in the risk preference (all $p<0.01$, two-sided two-sample $t$-tests).\footnote{We employ two-sided tests throughout all analyses. Unless otherwise specified, two-sample $t$-tests are used in this paper. Meanwhile, we conduct multiple hypotheses testing using the false discovery rate (FDR) method \citep{benjamini1995controlling} in 8 tests (4 models $\times$ 2 preferences), and report corrected $p$ values. This approach is also applied in the comparison in decisions between LLMs and humans in Case Study~2, where we regard the 20 tests (4 models $\times$ 5 scenarios) as a family for the correction.}

To investigate whether the responses of LLMs can be distinguished from those of humans, we conduct the Turing test \citep{mei2024turing,xie2024different}. In each comparison, we randomly draw one CCEI score from the LLM's distribution and one from the human distribution, and evaluate whether these two scores are equally likely to be generated under the human distribution and, if not, which one is more likely to be generated under the human distribution. We repeat this process 10,000 times to compute the corresponding probabilities. The results show that all four LLMs pass the Turing test, as they are estimated more likely or equally likely to be human in over half of the comparisons (Figure~\ref{fig:rationality—turing}). However, our randomly simulated agents fail the Turing test.

\subsubsection{Rationality in Experimental Conditions}
%Evaluating the impact of a single variation may require a set of experimental conditions. 
To quantify to what extent LLMs respond significantly to the experimental variations, we first construct an overall sensitivity score  as follows:
\[
\lambda(\mathcal{S})= \Sigma_{i \in \mathcal{I}, m \in \mathcal{M}, s \in \mathcal{S}} \frac{\mathbb{1} \{p_{ims} < 0.05\}}{|\mathcal{I}| |\mathcal{M}| |\mathcal{S}|}.
\]
In the equation, $\mathcal{I}$ is the set of four LLMs, $\mathcal{M}$ denotes the set of preference domains in Case Study 1 or decision scenarios in Case Study 2. $\mathcal{S}$ represents the set of experimental conditions in each variation. $p$ denotes the $p$ value from two-sample $t$-tests,
and the significance level is set at 0.05.\footnote{Notably, we correct the $p$ values by controlling FDR \mbox{\citep{benjamini1995controlling}}, where all tests within each variation are considered as a family for correction. In Case Study 1, we perform multiple hypothesis testing in 112 tests (4 models $\times$ 2 preferences $\times$ 14 conditions) for the persona; while for the dialogue type and answer type, similar analysis is conducted in 8 tests (4 models $\times$ 2 preferences). In Case Study 2, the number of tests within a family is 280 (4 models $\times$ 5 scenarios $\times$ 14 conditions) for the persona and 20 (4 models $\times$ 5 scenarios) for the answer type.} $\mathbb{1}(\cdot)$ is an indicator function.
To further measure the sensitivity score under each preference domain (decision scenario) $m\in\mathcal{M}$, we also calculate
\(
\lambda_m(\mathcal{S})= \Sigma_{i \in \mathcal{I},  s \in \mathcal{S}} \frac{\mathbb{1} \{p_{ims} < 0.05\}}{|\mathcal{I}| |\mathcal{S}|}.
\) By definition, both $\lambda$ and $\lambda_m$ belong to $[0, 1]$. The higher the values, the higher the sensitivity. 

\begin{figure}[htbp]
    \centering
    \includegraphics[width=1\linewidth]{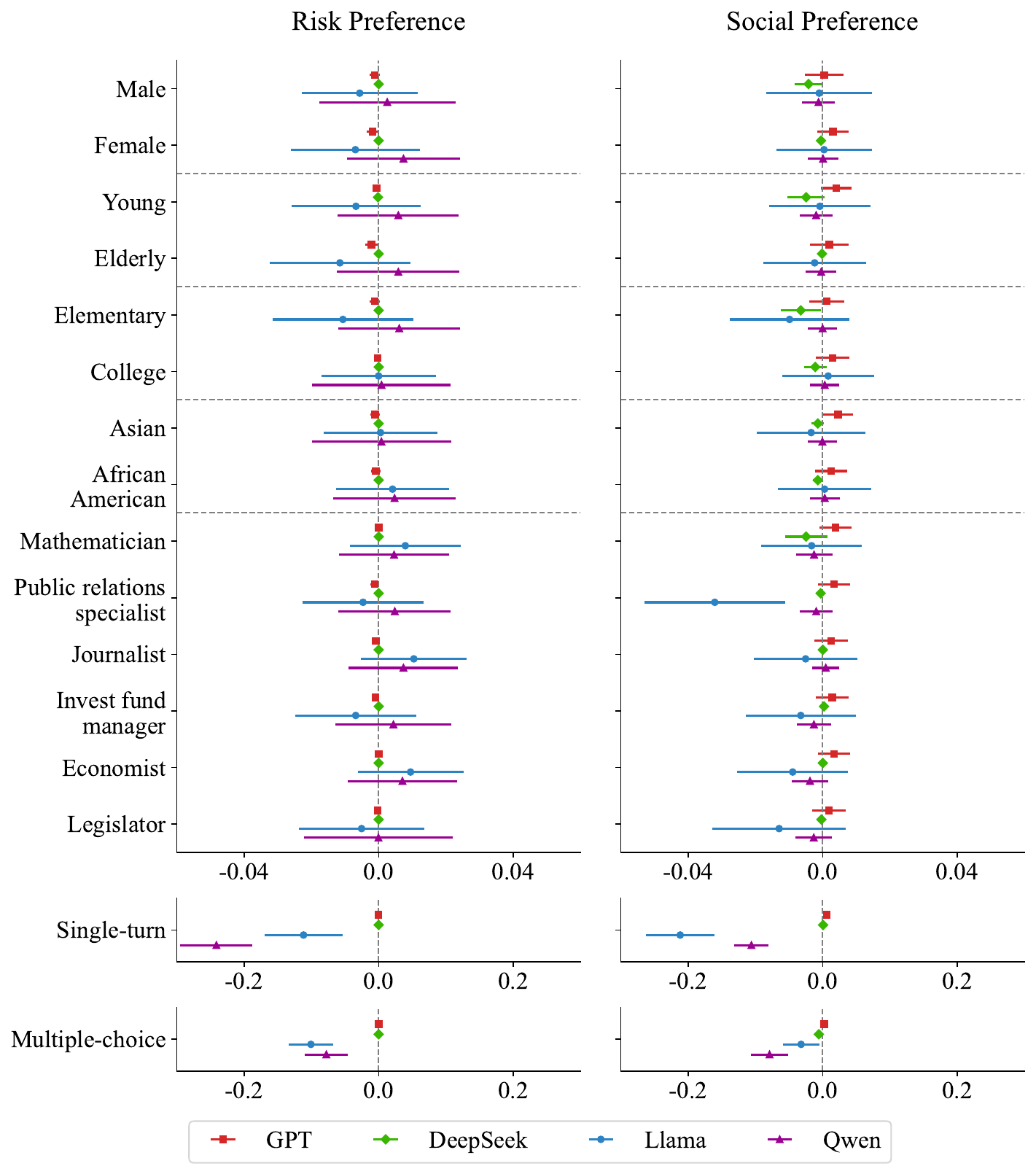}
    \caption{Comparison in rationality of LLMs. The markers represent differences in average CCEI scores of LLMs between experimental and baseline conditions. The error bars represent 95\% confidence intervals.}
    \label{fig:rationality-variation}
\end{figure}

\textit{Persona.} 
Using the persona specified in the system message, we calculate the average CCEI scores in different demographic conditions. Figure~\ref{fig:rationality-variation} presents the differences in CCEI between each experimental condition and the baseline condition. 
The results show that the rationality of all four models is not significantly influenced by the persona (all $p > 0.1$), leading to a sensitivity score $\lambda(\mathcal{S})$ of 0.
In general, the findings suggest that the persona has no impact on the economic rationality of LLMs.

\textit{Single-turn vs. multi-turn dialogue.} 
Compared to the baseline condition, the single-turn dialogue approach does not substantially affect the CCEI scores of GPT (risk: 0.999 vs. 1.000, $p > 0.1$, social: 1.000 vs. 0.994, $p < 0.01$) and DeepSeek (risk: 1.000 vs. 1.000, social: 1.000 vs. 0.999, $p < 0.01$) (Figure~\ref{fig:rationality-variation}). However, when using the single-turn dialogue, the CCEI of Llama decreases from 0.953 to 0.841 in the risk preference and from 0.968 to 0.756 in the social preference; that of Qwen decreases from 0.980 to 0.739 in the risk preference and from 0.994 to 0.889 in the social preference (all $p < 0.01$).
As a result, $\lambda(\mathcal{S})$ equals 75.0\%. These declines may arise from the limitation of single-turn dialogue in supporting sequential reasoning processes.

\textit{Open-ended vs. multiple-choice answer type.} 
As shown in Figure~\ref{fig:rationality-variation}, compared to the baseline condition using the open-ended answer type, the multiple-choice answer type does not affect the CCEI scores for GPT (risk: 1.000 vs. 1.000, social: 0.996 vs. 0.994, $p>0.1$) and DeepSeek (risk: 1.000 vs. 1.000, social: 0.994 vs. 0.999, $p = 0.060$). 
In contrast, under the multiple-choice answer type, the average CCEI scores in both domains decrease significantly for Llama (risk: 0.853 vs. 0.953, $p<0.01$, social: 0.936 vs. 0.968, $p=0.033$) and Qwen (risk: 0.902 vs. 0.980, social: 0.916 vs. 0.994, both $p<0.01$). This decrease mirrors the pattern observed in GPT-3.5-Turbo \citep{chen2023emergence}. Although the advancement from GPT-3.5-Turbo to GPT-4o over the past two years has eliminated the negative impact of discrete choices, this issue persists for smaller models such as Llama and Qwen.

\textit{Other variations.} 
In addition to the three variations discussed above, we also examine the influences of other factors in Supplementary Section~\ref{app:case1-other variation}. %Appendix~\ref{app:case1}
In summary, temperature settings between 0 and 1 do not affect the rationality of any of the four LLMs. 
The inclusion of examples does not affect the rationality in the risk preference, but does affect in the social preference. The existence of hypothetical incentives has no substantial impact on the rationality of four LLMs, while the rationality of Qwen surprisingly decreases as stake sizes increase in both preferences.

\section{Case Study 2: LLMs’ Economic Preferences}

In this section, we repeat our exercise using another set of tasks in Mei et al. \citep{mei2024turing}, 
which complement the economic rationality measured in Case Study 1. 

\subsection{Experimental Design}

We select the same four LLMs as in Case Study 1 and interact with them through the API using Python. The evaluated tasks include the dictator game, the ultimatum game, the public goods game, and the bomb risk game. In each game, the LLMs are required to make one decision, with the following exception---in the ultimatum game, the LLMs separately play the role of proposer, who suggests an allocation, and responder, who determines the minimum proposal they are willing to accept. 
As a result, we collect the decisions of LLMs across five distinct decision scenarios from the four games. A detailed description of these tasks can be found in Mei et al. \citep{mei2024turing}.
To avoid learning effects and simplify the analysis, we focus solely on the first round of decisions in each scenario. LLMs' choices are required to be highlighted with ``[[]]''. Data collection continues until 100 valid responses are obtained per scenario.

In the baseline condition, we adhere to the methodology in Case Study 1. Building upon this baseline, we implement two systematic variations concerning the persona and answer type. We do not examine the variation of dialogue type, as single-turn dialogue type is not feasible in these games. The prompts are detailed in Supplementary Section~\ref{prompt:case2}.

\begin{enumerate}[itemsep=3pt,parsep=0pt,topsep=3pt,partopsep=3pt]
    \item \textit{Persona.} We independently vary gender, age, education level, and ethnicity following Chen et al. \citep{chen2023emergence}, as well as occupations based on Mei et al. \citep{mei2024turing}.    
    \item \textit{Answer type.} Given that the feasible choice sets in all scenarios, except the public goods game, range from 0 to 100, we discretize this range into 21 options with a gap of 5. 
    In the public goods game, where the feasible choice set ranges from 0 to 20, we maintain the same number of 21 options and reduce the gap to 1.
\end{enumerate}

\subsection{Results for LLMs' Preferences}

\subsubsection{Preferences in Baseline Condition}\label{sec:pref-base}

\begin{figure}[htbp]

\begin{subfigure}{1\textwidth}
    \centering
    \includegraphics[width=1\linewidth]{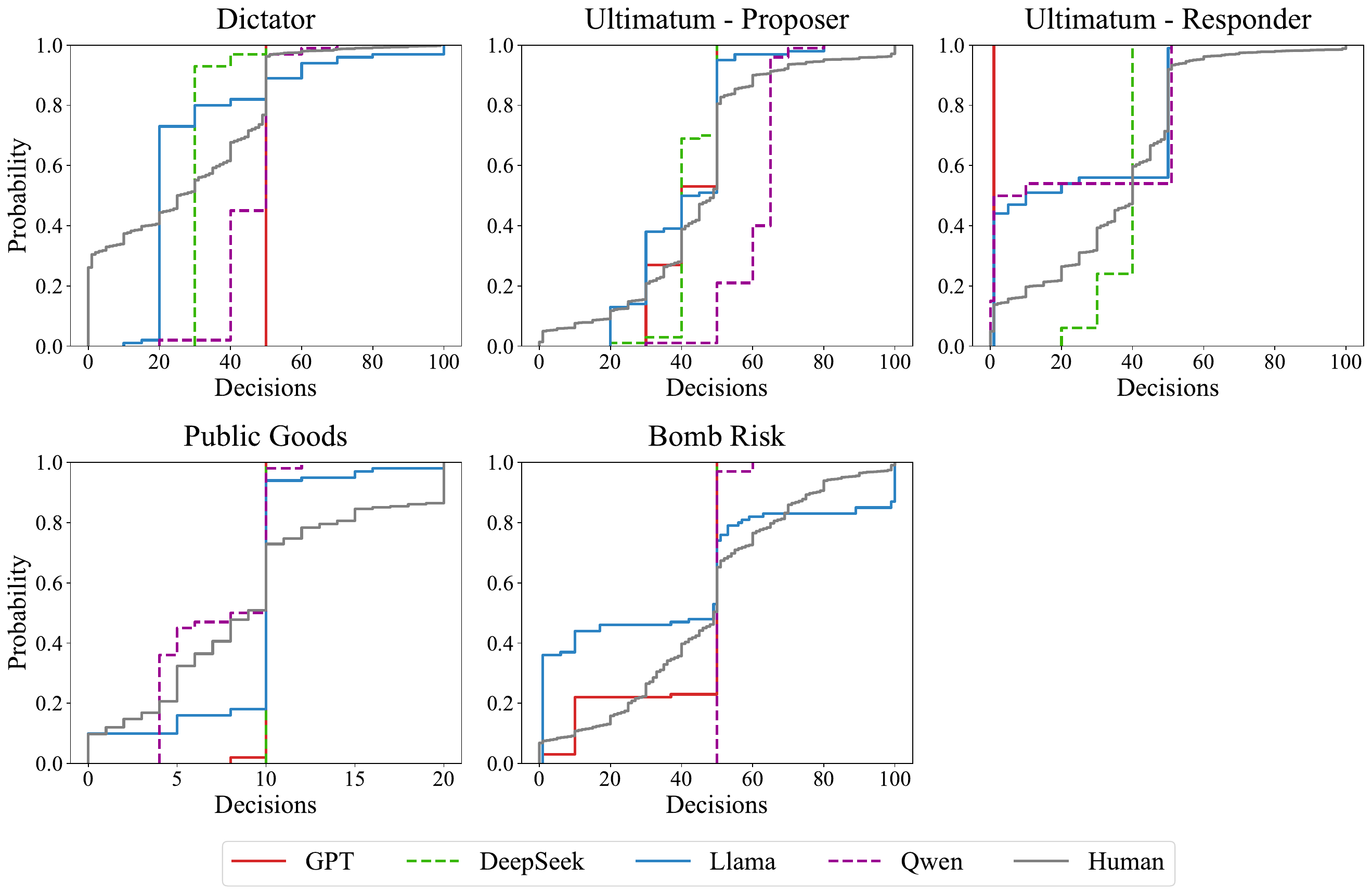}
    \caption{Cumulative Distribution of Decisions}
    \label{fig:decision—base}
\end{subfigure}

\begin{subfigure}{1\textwidth}
    \centering
    \includegraphics[width=1\linewidth]{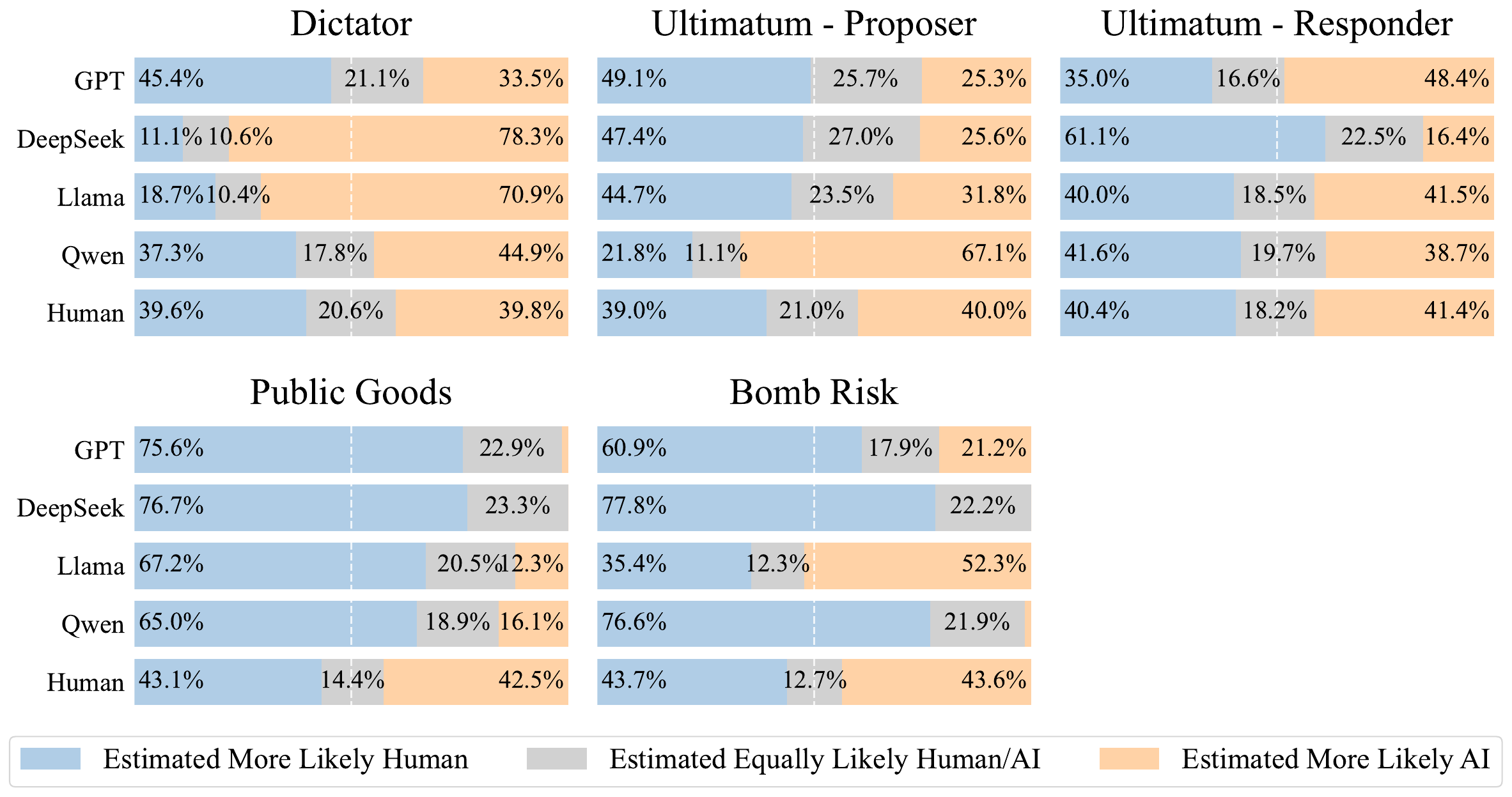}
    \caption{Turing test of Decisions}
    \label{fig:decision—turing}
\end{subfigure}
\caption{Baseline results of decisions. Human data are from Mei et al. \citep{mei2024turing}. In Panel b, proportions below 5\% are not labeled with numbers. The white dashed lines represent the probability of 50\%. In the human condition, both CCEI scores in each comparison are randomly drawn from the human distribution.}
    \label{fig:case2_base}
\end{figure}

Figure~\ref{fig:decision—base} describes the cumulative distribution of LLMs' decisions in five scenarios. Human data from Mei et al. \citep{mei2024turing} are also included as a reference. In the following, we compare the decisions between LLMs and human subjects for each scenario.

\begin{enumerate}[itemsep=3pt,parsep=0pt,topsep=3pt,partopsep=3pt]
    \item \textit{Dictator game.} GPT, DeepSeek, and Qwen allocate more endowment to the other player in the dictator game, compared to humans (all $p<0.05$).
    \item \textit{Proposer in ultimatum game.} Compared to human subjects, Qwen allocates more as the proposer in the ultimatum game ($p<0.01$).
    \item \textit{Responder in ultimatum game.} All LLMs, except DeepSeek, are willing to accept less money as responders in the ultimatum game compared to human subjects (all $p < 0.01$). Notably, GPT consistently accepts as low as \$1 across all 100 simulations.
    \item \textit{Public goods game.} DeepSeek contributes more than human subjects in the public goods game, while Qwen contributes less (both $p < 0.01$). 
    \item \textit{Bomb risk game.} GPT and Llama are more risk averse than human players; conversely, DeepSeek and Qwen dominantly choose to open 50 boxes---half of the total boxes---exhibiting less risk aversion than human subjects (all $p<0.05$).

\end{enumerate}

Together, the four LLMs exhibit divergent preferences within each scenario, implying considerable heterogeneity in decisions between models. 
Moreover, Figure \ref{fig:decision—base} shows that decisions from open-sourced models with smaller parameter sizes, such as Llama and Qwen, exhibit greater dispersion, which suggests higher heterogeneity in the inferred preferences of these models. In contrast, the behavior of GPT and DeepSeek is more consistent.\footnote{For example, GPT consistently selects \$50 in the dictator game and sets the acceptance threshold at \$1 when acting as the responder in the ultimatum game. DeepSeek consistently contributes half of its endowment in the public goods game and opens half of the boxes in the bomb risk game. In a quantitative analysis, we calculate the normalized standard deviation of each LLM using $ \tilde{\sigma}_i =  \Sigma_{m\in \mathcal{M}} \frac{\sigma_{im}}{L_{m} |\mathcal{M}| }$. Here, $\sigma_{im}$ denotes the standard deviation of LLM $i$ in decision scenario $m \in \mathcal{M}$ in the baseline condition, and $L_m$ represents the length of the feasible set interval of scenario $m$, which is 20 for the public goods game and 100 for the other scenarios. The normalized standard deviations $\tilde{\sigma}_i$ for GPT, DeepSeek, Llama, and Qwen are 0.054, 0.030, 0.218, and 0.110, respectively. For comparison, the corresponding value for human decisions is 0.231.} 

Moreover, GPT passes the Turing test in all games, whereas DeepSeek and Qwen each fail in one of the games, and Llama fails in two of them (Figure~\ref{fig:decision—turing}). These results suggest that more advanced LLMs are capable of exhibiting human-like behavior, while less capable models fall short of this standard.

\subsubsection{Preferences in Experimental Conditions}

\textit{Persona.} 
Two key observations emerge. First, the decisions of LLMs in behavioral games are more sensitive to personas than in budgetary tasks ($\lambda(\mathcal{S})$: 35.7\% vs. 0\%, $p < 0.01$, proportion test).
Moreover, their choices in the public goods game are least likely to be influenced by personas among all games, with $\lambda_m(\mathcal{S})$ equal to 12.5\%. Conversely, the sensitivity score increases significantly and reaches the highest when LLMs act as responders in the ultimatum game ($\lambda_m(\mathcal{S})$: 51.8\% vs. 12.5\%, $p<0.01$, proportion test). These findings suggest that the impact of personas depends on the nature of the tasks evaluated.

Second, LLMs' outputs are more frequently influenced by occupational personas than by other demographic characteristics ($\lambda(\mathcal{S})$: 48.3\% vs. 26.3\%, $p<0.01$, proportion test). 
This underscores that the effect of personas is dependent on the specific types of personas.
Taken together, these observations suggest that we cannot draw definitive conclusions about how personas influence experimental results, as their impact varies across tasks and persona types.
% \textcolor{blue}{These findings provide evidence on against assigning specific personas in prompts, as the results may not be robust. --Move to the tactic part.}

\textit{Open-ended vs. multiple-choice answer type.} 
Compared to the open-ended answer type in the baseline condition, responding within a restricted set of options significantly alters LLMs' outputs. Specifically, $\lambda(\mathcal{S})$ equals 55.0\% under the multiple-choice answer type, indicating that the answer type significantly influences more than half of LLMs' decisions. Moreover, this value is marginally higher than that obtained under persona variations ($\lambda(\mathcal{S})$: 55.0\% vs. 35.7\%, $p=0.084$, proportion test), demonstrating that LLMs' decisions are more sensitive to the answer type than personas.

\begin{figure}[htbp]
    \centering
    \includegraphics[width=1\linewidth]{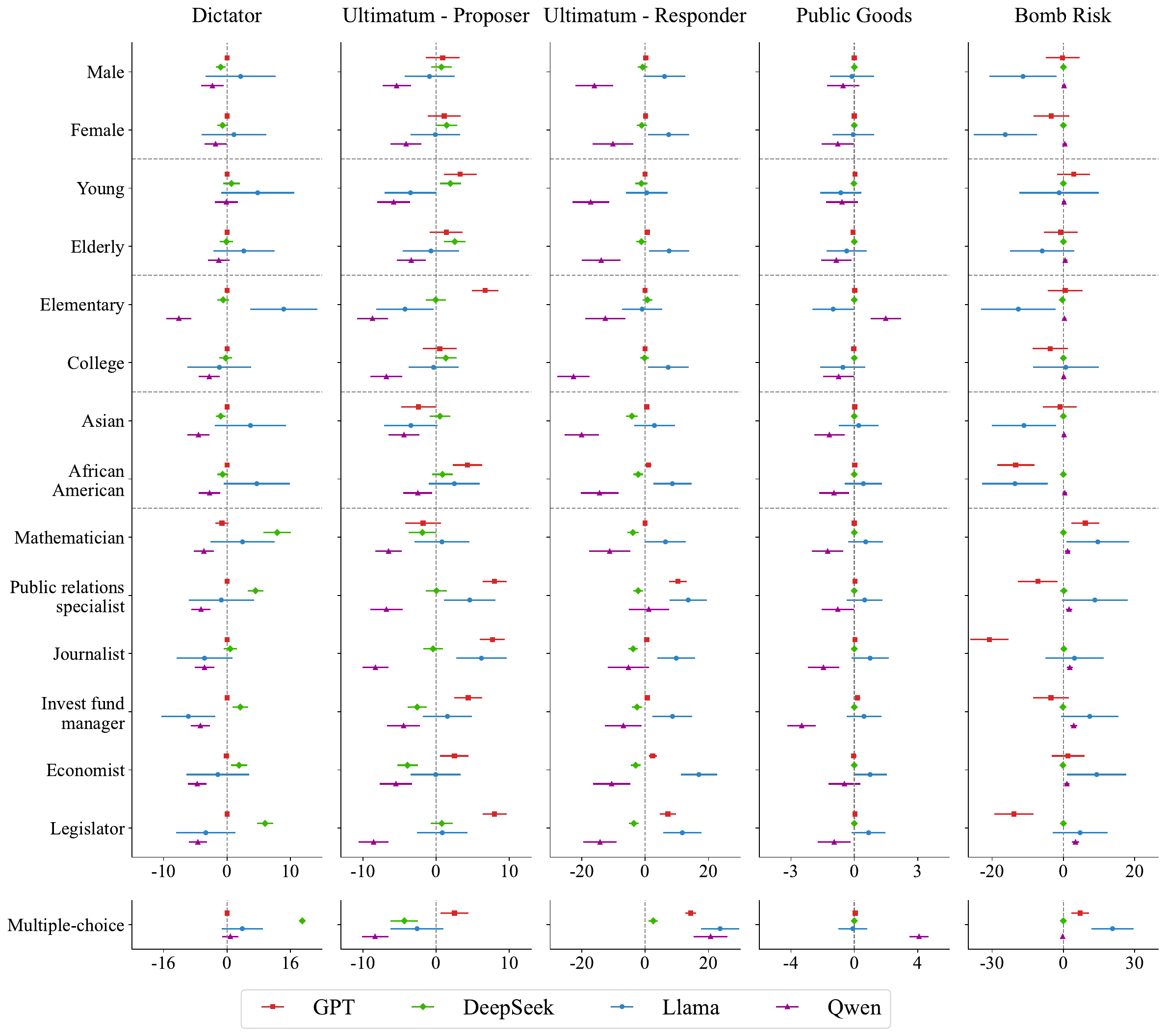}
    \caption{Comparison in decisions of LLMs. The markers represent differences in the mean value of LLMs' decisions between experimental and baseline conditions. The error bars represent 95\% confidence intervals. }
    \label{fig:case2_variation}
\end{figure}

\section{Tactics for LLM Experiments}
\label{sec:tactic}

Drawing from key principles of human experiments, protocols in prior LLM research, and our own experimental findings, we present seven key tactics for conducting LLM experiments. These tactics serve two main purposes.
First, they aim to align the procedures in LLM experiments with those in human experiments as closely as possible, thus facilitating the comparison in experimental results between LLM agents and human subjects. 
Second, with the burgeoning of LLM experiments, we hope these tips can be useful in improving the replicability of LLM experimental results.
In the following, we will discuss each tactic by integrating insights from both experimental methods in economics and prompt engineering in artificial intelligence.

\subsection{Experimental Design}

\textit{Temperature.} 
The selection of temperature involves the trade-off between adopting the solution with higher probability and exploring a broader solution space \citep{renze2024effect}. Human participants often exhibit a certain degree of heterogeneity. To mimic their behavior, we should allow the variation of responses between LLM agents. Meanwhile, both existing studies \citep{chen2023emergence,kosinski2024evaluating,peeperkorn2024is,renze2024effect} and our results (Supplementary Figure~\ref{fig:rationality_app}) have shown that there are no significant behavioral differences in LLMs' outputs when different temperatures are used. Therefore, we suggest using the default temperatures recommended by model providers. When there are no default values, setting the temperature to 1 is a natural choice. This value does not scale the logits (see Supplementary Section~\ref{app:how_to_prompt}), thus best mirroring the application of softmax during the training process, which does not involve the temperature parameter \citep{radford2018improving}. In addition, some model providers allow users to specify a random seed parameter. To enhance the replicability of results, researchers can set the seed when prompting LLMs via the API.
We summarize these discussions in the following. 

\textit{\textbf{Tactic 1}: Use the default temperature and fix the random seed.} %Use a fixed random seed to enhance replicability

\textit{Persona.} 
Following the conventions of experimental economics, where experimenters do not explicitly recruit participants with specific demographics unless this is part of the research design, we recommend not including specific personas in the prompts. Our two case studies further support this tactic that the use of personas may introduce uncontrolled variables without consistently yielding significant or generalizable effects.
In general, LLMs should be regarded as counterparts to representative individuals from the general human population. However, if the goal is to assess whether LLMs with diverse demographic characteristics respond differently to tasks \citep{chen2023emergence,mei2024turing,wang2025large}, we can implement it by assigning different personas. The caveat here is that the comparison in results between such studies should be cautious due to the differences in selected tasks, evaluated models, and persona types.

Moreover, LLMs may fail to understand the meaning of assigned personas and under-represent the heterogeneity across different subgroups. For example, assigning male or female identities may cause LLMs' responses to shift in the same direction, compared to using no personas at all \citep{deng2025can}. 
LLMs with a persona also tend to resemble individuals without that identity describing those who possess it, rather than representing individuals with that identity. This probably comes from the nature of the training data: The demographic characteristics of the authors who generate these data are rarely disclosed. When such identities appear in the corpus, they are usually mentioned by others rather than self-identified by the individuals themselves \citep{wang2025large}.

\textit{\textbf{Tactic 2}: Avoid assigning personas in prompts unless research requires.}

\textit{Incentive.} 
Based on the induced value theory \citep{smith1976experimental,smith1982microeconomic}, the use of monetary incentives has become a standard practice in laboratory experiments, although the extent to which human behavior differs under real versus hypothetical incentives remains inconclusive \citep[e.g.,][]{holt2002risk, laury2009insurance,branas2023paid,gneezy2024real}. 
Since real monetary rewards cannot be actually applied to LLMs, the feasible alternative is to instruct LLMs to maximize their hypothetical payoffs. Therefore, clear payment rules could be included in the prompts as in the instructions for human participants. However, unlike humans, LLMs may lack an intrinsic utility function; their data generation is driven by externally imposed optimization rules in training processes. This indicates that LLMs are less likely to report untruthfully, the issue that incentives are designed to address. Note that although LLMs may exhibit hallucination characterized by nonsensical or unfaithful responses, this phenomenon results from the training data, training processes, and decoding strategies \citep{huang2025survey} rather than strategic deception. Based on these facts and the experimental results (Supplementary Figure~\ref{fig:rationality_app}), we conjecture that LLMs' performance may change little when such hypothetical incentives are removed.

\textit{\textbf{Tactic 3}: Use the typical payment scheme to be comparable to human experiments.}

\subsection{Experimental Implementation}

\textit{Illustrative example and understanding question.} 
In human experiments, particularly those using complex tasks, illustrative examples and understanding questions are routinely incorporated to facilitate comprehension. 
We apply the same principles for LLM experiments. 

\textit{\textbf{Tactic 4}: Incorporate examples and comprehension questions especially for complex tasks.}

\textit{Dialogue type.} 
Since the structure of multi-turn dialogue is more similar to the sequential decision-making process of human subjects, we suggest using it for tasks involving multiple questions for a single agent.
In contrast, although single-turn dialogue offers advantages in the efficiency of model inputs, it may constrain the economic rationality of certain models, particularly those with smaller parameter sizes (Figure~\ref{fig:rationality-variation}).

\textit{\textbf{Tactic 5}: 
Use multi-turn dialogue for tasks with multiple questions.}

\textit{Answer type.} 
The construction of the answer format can influence model performance. For example, providing Likert scales or multiple choices may activate memory in the model training corpus, or enable models to infer experimental purposes and act in a demanding way \citep{kosinski2024evaluating}. Furthermore, discretizing continuous feasible sets has been shown to significantly decrease the performance of models in both Chen et al. \citep{chen2023emergence} and our Case Study 1 (Figure~\ref{fig:rationality-variation}). Consequently, the open-ended answer format is recommended rather than multiple-choice, unless the latter format is required by the nature of tasks. For example, the Linda problem requires multiple options with inclusion relationships to measure the conjunction fallacy \citep{kahneman1973psychology}. 

\textit{\textbf{Tactic 6}: Use an open-ended answer format whenever feasible.}

\textit{Invalid answer.} 
In prior LLM experiments, the common practice is to exclude invalid answers from the dataset and continue data collection until the target sample size is reached \citep[e.g.,][]{chen2023emergence,brookins2024playing,goli2024frontiers,mei2024turing}. However, the frequency of invalid answers also provides valuable information on model performance \citep{macmillan2024ir,chen2025manager}. Therefore, although the content of invalid answers may remain unused, we believe that the proportion of such responses should be documented and reported as a performance metric.

\textit{\textbf{Tactic 7}: Report the proportion of invalid responses.}

\section{Discussion}

Integrating established principles from experimental economics with insights from artificial intelligence, this paper systematically investigates how key considerations from experimental design to implementation influence LLMs' outputs.
We find that assigned personas have a considerable impact on economic preferences but do not affect economic rationality. Furthermore,  the single-turn dialogue type reduces the economic rationality of two open-source models with small parameter sizes, Llama and Qwen, but not for the others, GPT and DeepSeek. Similarly, discretizing the continuous choice set into multiple choices leads to a reduction in rationality for Llama and Qwen, as well as substantial changes in decisions in behavioral games for all LLMs. These highlight the sensitivity of LLMs' performance to variations in experimental design and implementation. Based on these observations, we summarize seven tactics for designing and conducting LLM experiments, thereby contributing to more comparable and replicable research practices.

Our paper adds to the emerging literature at the intersection of LLMs and experimental economics, with a particular focus on advancing methodological approaches. Previous research mainly explores how LLMs can be effectively used as supplementary tools at various stages of human experiments \citep{charness2023generation}, and how to use generative AI in experimental research \citep{chang202412}. In addition to these studies, we shift the perspective by treating LLMs as experimental objects and develop a framework for standardizing experimental protocols in LLM experiments. Moreover, this framework is grounded in two case studies, which further enhances its validity.

Our study also contributes to improving the replicability and generalizability of LLM experiments. Several studies have assessed the replicability of experimental studies with human subjects \citep{camerer2016evaluating,camerer2018evaluating,holzmeister2025examining}. The findings indicate that experiments in economics exhibit a higher rate of replicability compared to those in psychological sciences \citep{camerer2016evaluating}, which may be attributed to the rigorous methodological standards and established norms developed over decades within experimental economics \citep{croson2005method,camerer2016evaluating,frechette2022experimental,niederle2025experiments}. As LLM experiments are still in their early stages, similarly standardized protocols have yet to be formalized. Our study addresses this gap by proposing protocols that aim to increase the methodological rigor. In addition, aligning these protocols with those used in human experiments can improve the generalizability of LLM-based findings to human experiments, offering new insights into human behavior.

Finally, we acknowledge several limitations and outline directions for future research emerging from this work. First, we focus exclusively on the individual decision making of LLMs. Extending this framework to accommodate multi-agent settings and human–AI hybrid experiments \citep{engel2024integrating} will be important for broadening its applicability. Second, our study proposes a set of protocols and evaluates them in the context of text-based interactions. As LLMs' architectures continue to evolve, especially with the development of multimodal systems that integrate text, voice, image, and video data, further validation will be necessary to ensure the effectiveness of these protocols for next-generation models. Third, while we assess how design factors influence LLMs' performance, the underlying mechanisms are not understood sufficiently. Enhancing the interpretability of LLMs' decision-making is crucial to validate whether observed behavior reflects genuine human-like reasoning or simply mimics patterns in training data \citep{olah2020zoom,bommasani2021opportunities,black2022interpreting}. Addressing these issues will improve experimental protocols and observations, develop interpretations and theories of LLMs' behavior, and support the effective application and integration of LLMs into everyday life.

\onehalfspacing
\appendix

\newpage

\begin{center}
    \LARGE{\textbf{Online Appendices}}
\end{center}

\startcontents[sections]
\printcontents[sections]{l}{1}{\setcounter{tocdepth}{2}}

\newpage

\setcounter{table}{0}
\renewcommand{\thetable}{\arabic{table}} 
\setcounter{figure}{0}
\renewcommand{\thefigure}{\arabic{figure}}

\begin{landscape}
\thispagestyle{plain}
    \begin{table}[htbp]
\renewcommand{\tablename}{Supplementary Table}
  \centering
  \caption{The Description of Representative LLM Experiments}
  \label{tab:llms_exp_sum}%
  \footnotesize
    \begin{threeparttable}[b]
\begin{tabular}{p{10em}p{11em}p{21em}p{20em}} 
\toprule
\makecell[c]{Study} & \makecell[c]{Journal/Conference} & \makecell[c]{Model} & \makecell[c]{Representative task} \\ 
\midrule
 \citet{binz2023using} & PNAS & GPT-3 (ada, babbage, curie, davinci) & Linda problem and CRT \\ 
\midrule
\citet{chen2023emergence} & PNAS & GPT-3.5-Turbo & Budgetary tasks \\ 
\midrule
\citet{hagendorff2023human} & Nature Computational Science & GPT-1, GPT-2XL, GPT-3 (ada, babbage, curie, davinci-001, davinci-002, davinci-003), ChatGPT-3.5, ChatGPT-4 & Semantic illusions and CRT \\
\midrule
\citet{webb2023emergent} & Nature Human Behaviour & GPT-3 (text-davinci-003, code-davinci-002, text-davinci-002, davinci), GPT-4 & Raven’s  Progressive Matrices \\ 
\midrule
\citet{brookins2024playing} & Economics Bulletin & GPT-3.5-Turbo & Dictator game and prisoner's dilemma game \\ 
\midrule
\citet{suri2024large} & Journal of Experimental Psychology: General & GPT-3.5, ChatGPT-4, Bard & Linda problem \\ 
\midrule
\citet{goli2024frontiers} & Marketing Science & GPT-3.5-Turbo, GPT-4 & Intertemporal choice task \\ 
\midrule
\citet{macmillan2024ir} & Royal Society Open Science & GPT-3.5, GPT-4,  Bard, Claude 2, Llama 2 (7B, 13B, 70B) & Linda problem and Wason task \\ 
\midrule
\citet{mei2024turing} & PNAS & GPT-3.5-Turbo, GPT-4, ChatGPT (subscription-based Web version (Plus), freely available Web version (Free)) & Dictator game, ultimatum game, trust game, bomb risk game, public goods game, prisoner's dilemma game \\ 
\midrule
\citet{kosinski2024evaluating} & PNAS & GPT-1, GPT-2XL, GPT-3 (ada, babbage, curie, davinci-001, davinci-002, davinci-003), BLOOM, ChatGPT-3.5-turbo, ChatGPT-4 & Unexpected contents task and unexpected transfer task \\ 
\midrule
\citet{wangwill2024} & COLM & GPT-3.5-Turbo, GPT-4, PaLM 2, Llama 2 70B & Linda problem \\ 
\midrule
\citet{chen2025manager} & Manufacturing \& Service Operations Management & GPT-3.5-Turbo, GPT-4 & Linda problem and CRT \\
\bottomrule
\end{tabular}
    \vspace{0.3em}
    \begin{minipage}{\linewidth}
	\footnotesize \textit{Notes}: PNAS: \textit{Proceedings of the National Academy of Sciences}. COLM: \textit{Conference on Language Modeling}. CRT: Cognitive reflection task.
\end{minipage}
\end{threeparttable}
\end{table}

\end{landscape}

\begin{landscape}
\thispagestyle{plain}
\begin{table}[htbp]
\renewcommand{\tablename}{Supplementary Table}
  \centering
  \caption{The Prompt Parameter Specification of Representative LLM Experiments}
  \label{tab:llms_exp_eva}%
\begin{threeparttable}[b]
\begin{tabular}{lcccccccc}
\toprule
\makecell[c]{Study} & Temperature & Persona & Incentive & Example & \makecell[c]{Understanding \\ question} & \makecell[c]{Dialogue \\ type} & \makecell[c]{Answer \\ type} & \makecell[c]{Invalid \\ answers} \\
\midrule
\citet{binz2023using} & 0 &  & Both &  &  & Multi-turn & Both &  \\
\citet{chen2023emergence} & 0, 0.5, 1 & Both & Y &  & Y & Single-turn & Both &  \\
\citet{hagendorff2023human} & 0 &  &  &  &  & $-$ & Open & Y \\
\citet{webb2023emergent} & 0 &  &  &  &  & Multi-turn & Both &  \\ 
\citet{brookins2024playing} & 1 & Y & Y &  &  & $-$ & Both & Y \\ 
\citet{suri2024large} & $-$ &  &  &  &  & $-$ & Both &  \\ 
\citet{goli2024frontiers} & 1 &  & Y &  &  & $-$ & Choice &  \\ 
\citet{macmillan2024ir} & 1 &  &  &  &  & $-$ & Both & Y \\ 
\citet{mei2024turing} & 1 & Both & Y & Both &  & Multi-turn & Both &  \\ 
\citet{kosinski2024evaluating} & 0, 1 &  &  &  &  & $-$ & Open &  \\
\citet{wangwill2024} & 0.7 &  &  & Both &  & $-$ & Both &  \\
\citet{chen2025manager} & 1 &  & Both &  &  & $-$ & Both & Y  \\
\bottomrule
\end{tabular}
    \vspace{0.3em}
\begin{minipage}{\linewidth}
	\footnotesize  \textit{Notes}: In terms of temperature, we only summarize the temperature of models in GPT family for comparison, and ``$-$'' represents the parameter is not revealed in the study.
    ``Y'' represents the parameter or feature is specified in the study, while the black represents the opposite. ``Both'' represents that the study includes the parameter or feature in some experimental conditions or tasks and not in the others. For dialogue type, ``$-$'' represents the feature is not applicable, as each LLM agent only needs to answer one question. For answer type, ``Open'' represents the open-ended type and ``Choice'' represents the multiple-choice type. ``Invalid answers'' describes whether the study reports the proportion of invalid answers.
\end{minipage}
\end{threeparttable}
\end{table}
\end{landscape}

\newpage

\begin{figure}[H]
\renewcommand{\figurename}{Supplementary Figure} 
	\centering
	\begin{subfigure}{0.24\textwidth}
		\centering
		\includegraphics[height=0.27\textheight]{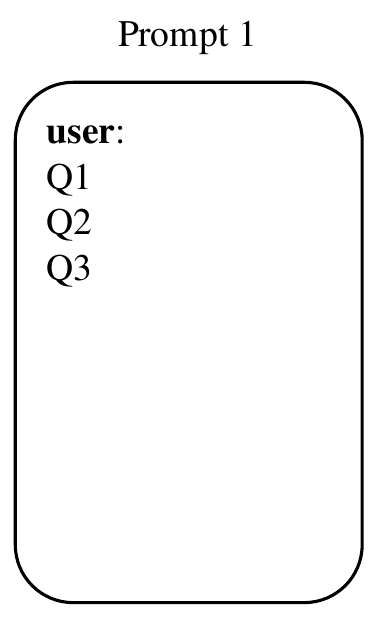}
        \caption{Single-turn dialogue}
        \label{fig:number of turns-single}
    \end{subfigure} % do not put space
    \begin{subfigure}{0.72\textwidth}
		\centering
		\includegraphics[height=0.27\textheight]{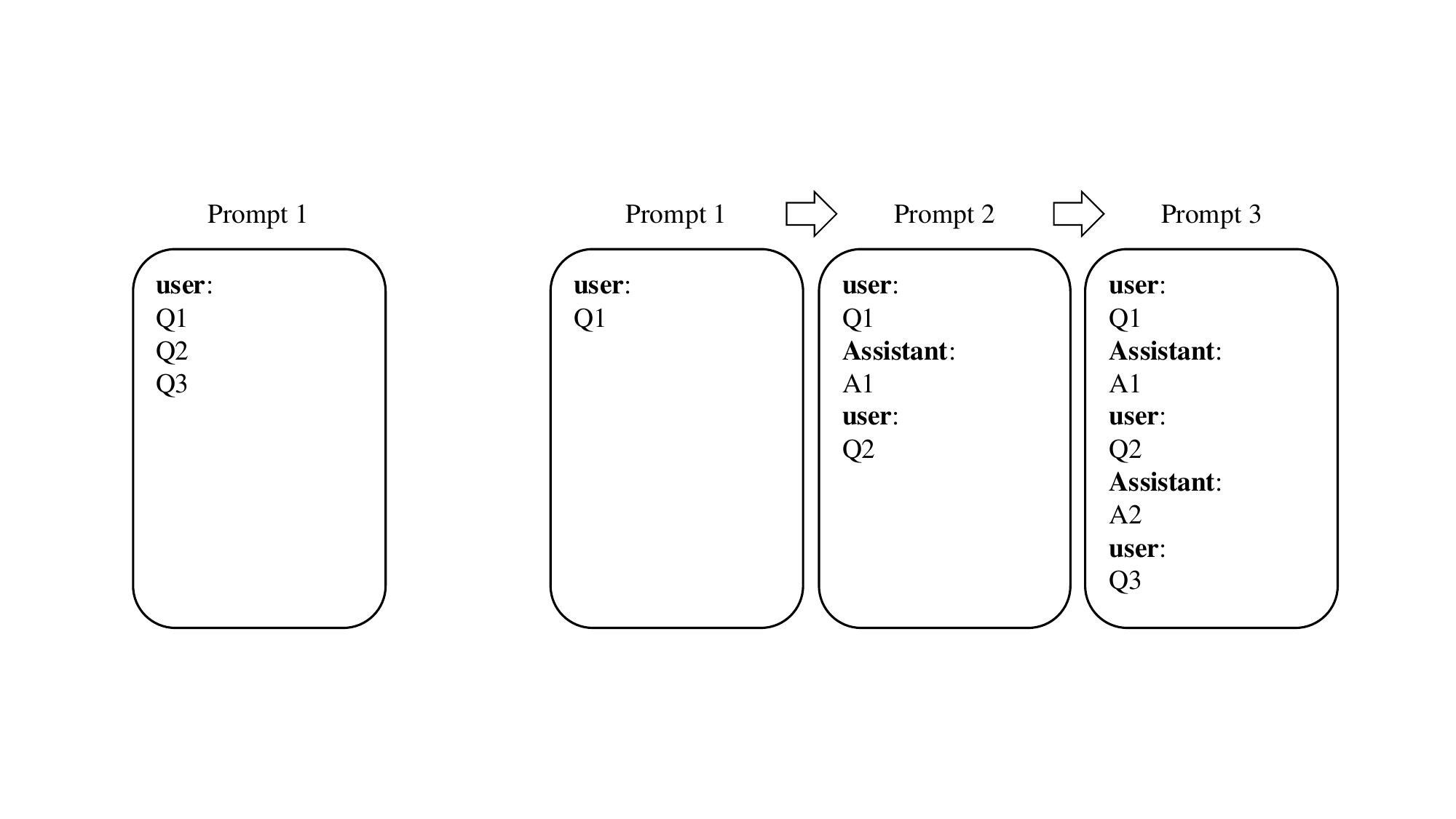}
        \caption{Multi-turn dialogue}
        \label{fig:number of turns-multi}
    \end{subfigure}\
	
    \caption{Illustration of Dialogue Type}
    \label{fig:number of turns}

\end{figure}

\newpage

\section{Principles in Human Experiments}\label{sec:human exp}

\subsection{Key Principles}

To review key principles for human experiments, we resort to classical textbooks and survey papers that offer comprehensive discussions on experimental methodology. We choose two textbooks \citep{davis1993experimental,friedman1994experimental}, which have high citation counts.\footnote{As of April 9, 2025, the Google Scholar citation counts for \citet{davis1993experimental} and \citet{friedman1994experimental} were 3,197 and 1,442, respectively.} Croson \citep{croson2005method} is also included for its detailed comparison in experimental methods between economics and psychology. Two students who have taken the Ph.D. level experimental economics class independently read these three references and summarize the key methods for experimental design and implementation. We take the union of these principles and then distill them based on two criteria: 1) they are mentioned in at least two of the three sources, and 2) they have potential applicability to LLM experiments. Table~\ref{tab:human_rules} provides a summary of key guidelines from these three sources.

\begin{table}[htbp]
\renewcommand{\tablename}{Supplementary Table} 
  \centering
  \caption{Key Principals for Human Experiments}
    \begin{tabular}{lccc}
\toprule
  & DH1993 \citep{davis1993experimental} & FS1994 \citep{friedman1994experimental}  & Croson2005  \citep{croson2005method} \\
\midrule
\multicolumn{4}{l}{\textit{Panel A:  Basic Rules in Experimental Design}} \\ 
\midrule
Follow the induced value theory & \checkmark & \checkmark & \checkmark \\
Avoid deception & \checkmark & \checkmark & \checkmark \\
Minimize experimenter demand effects & \checkmark & \checkmark & \checkmark \\
\midrule
\multicolumn{4}{l}{\textit{Panel B: Execution Rules in Experimental Implementation}} \\
\midrule
Use context-neutral instructions  & \checkmark & \checkmark & \checkmark \\
Make instructions comprehensible &  & \checkmark & \checkmark \\
Provide illustrative examples &  & \checkmark & \checkmark \\ 
Incorporate comprehension assessments  &  & \checkmark & \checkmark \\
Prevent inter-subject communication  &  & \checkmark & \checkmark \\
\bottomrule
    \end{tabular}
  \label{tab:human_rules}%
\end{table}%

First, in the experimental design stage, we focus on the following principles.
\begin{enumerate}[itemsep=3pt,parsep=0pt,topsep=3pt,partopsep=3pt]
\item \textit{Follow the induced value theory.} For economic experiments, a central consideration is the alignment of monetary incentives with induced value theory. Specifically, payment schemes must adhere to principles such as monotonicity, salience, and dominance \citep{smith1982microeconomic}. Induced value theory ensures a linkage between participants' payoff and their decisions, thereby eliciting truthful responses. In economic experiments, cash is typically used as the reward medium, instead of other methods, e.g., grades or extra credit points.\footnote{Grades or extra credit points may not be valued by every student, whereas cash is usually non-satiable \citep{croson2005method}.} Moreover, though negative payment is possible in theory, subjects' payments are always positive in practice.

\item \textit{Avoid deception}. Economic experiments strictly prohibit deception on participants, such as the inconsistency between the actual and prespecified payment rules as well as the provision of false feedback. 
Deception undermines the trust of participants in experimenters, thus potentially changing their behavior in future experiments and contaminating the subject pool.
For example, deception on the payoff rule will weaken the link between the payoff and the participants' decisions due to distrust, leading to false reports \citep{croson2005method}.

\item \textit{Minimize experimenter demand effects (EDE).} Experimenter demand effects (EDE) also pose a threat to the validity of experimental results.\footnote{Specifically, EDE occurs when participants discern the experimental objectives and correspondingly modify their behavior. For example, \citet{branas2007promoting} includes ``Note that he relies on you” within the instruction and finds that it leads to more generous behavior.} Using non-deceptive obfuscation to conceal experimental objectives in certain cases is emphasized as a means of reducing EDE \citep{zizzo2010experimenter}.

\end{enumerate}

In summary, all of these principles are designed to guarantee truthful reports from human subjects. 
However, their applicability to LLM experiments might be limited, as LLMs generate outputs by sampling from probability distributions learned during the training processes, rather than through subjective decision-making. 

Second, during the experimental implementation stage, most rules emphasize the clarity of the instruction to ensure participant comprehension. 

\begin{enumerate}[itemsep=3pt,parsep=0pt,topsep=3pt,partopsep=3pt]
    \item \textit{Use context-neutral instructions.} Non-neutral language may additionally introduce participants' preferences and attitudes as confounders, thereby compromising the reliability of experimental results. For example, as shown in the instructions from \citet{plott1978experimental} in the Section~\ref{app:instruction}, participants are simply informed that they are trading a generic commodity without reference to specific categories such as fruits or meat. This approach eliminates the possibility that participants' choices are influenced by personal preferences for particular goods.
    \item \textit{Make instructions comprehensible.} The instructions should be easily understood. Meanwhile, the payment rules, as well as the available choices and actions, should be clearly outlined in the instructions.
    \item \textit{Provide illustrative examples.} In the case of complex tasks, illustrative examples are often used to enhance comprehension. 
    \item \textit{Incorporate comprehension assessments.} Understanding questions serves as an effective method to assess the participants’ comprehension of experimental instructions. When the subject pool is sufficiently large, as in online experiments, individuals who fail these questions can be immediately excluded from experiments. In other cases, the accuracy rate on the understanding questions can be taken into account in the analysis to control for variations in comprehension.
    \item \textit{Prevent inter-subject communication.} Prohibiting communication among participants is standard practice at all stages of experiments, except that communication itself is a focus of the experimental design. Communication can lead participants to infer the experimental objectives by learning from others, potentially inducing EDE. Moreover, it compromises the independence of the observations, thus reducing the statistical power of the analysis.
    
\end{enumerate}

In the implementation stage, the guide of instructions is particularly relevant for LLM experiments, as the instructions provided for human participants closely parallel the instructions for LLMs.

\subsection{Sample Instructions} \label{app:instruction}
In this section, we will take the instructions in Plott and Smith \citep{plott1978experimental} as an illustrative example to explain the rules listed above. The instructions are also reprinted in Friedman \citep{friedman1994experimental} in the appendix. The whole original instructions include four parts: General instructions, specific instructions for sellers, specific instructions for buyers, and market organizations. For simplicity concern, we include the first two parts to illustrate. Additional explanations are added in square brackets with underlines.

``\textit{General}

\textit{This is an experiment in the economics of market decision making. Various research foundations have provided funds for the conduct of this research. The instructions are simple, and if you follow them carefully and make good decisions you might earn a considerable amount of money which will be paid to you in cash after the experiment. \note{Make instructions comprehensible. Follow the induced value theory.}
}

\textit{In this experiment we are going to simulate a market in which some of you will be buyers and some of you will be sellers in a sequence of market days or trading periods. Two kinds of sheets will now be distributed—information for buyers and information for sellers. The sheets are identified and numbered. The number is only for data-collecting purposes. If you have received sellers’ information, you will function only as a seller in this market. Similarly, if you have received buyers’ information, you will function only as a buyer in this market. The information you have received is for your own private use. Do not reveal it to anyone. \note{Prevent inter-subject communication.}
}

\textit{This is a one commodity market in which there is no product differentiation. \note{Use context-neutral instructions.} That is, each seller produces a product which is similar in all respects to the products offered by the other sellers. A seller is free to sell to any buyer or buyers. Likewise, a buyer may purchase from any seller or sellers.}

\textit{Specific Instructions for Sellers}

\textit{During each market period, you are free to produce and sell any of the amounts listed on your information sheet. Assume that you produce only for immediate sale—there are no inventories. The dollar amounts listed in column 2 on your information sheet are your costs of producing that quantity.}

\textit{Your payoffs are computed as follows: At the beginning of the experiment you will receive starting capital of \$0.30. If you are able to make any sales, you will receive the difference between your sales revenue and your cost. For example, if you were to sell two units at \$100 each, total revenue would be \$200. Suppose your information sheet indicated that the cost of producing two units was \$190. Your total profit would then be \$200$-$\$190 $=$ \$10 for the trading period. If you sold two units for less than \$190 you would incur a loss. Column 3 will be useful to a seller in deciding at any time during a given trading period whether to sell an additional unit. Suppose a seller has already sold one unit at a profit, and wants to know if he should sell a second unit. If the additional cost of producing the second unit is \$10, then he will lose money on that unit if he sells it at any price below \$10. \note{Provide illustrative examples.} Obviously, these figures are illustrative only and should not be assumed to apply to the actual sellers in this experiment. \note{Minimize experimenter demand effects.}}

\textit{
All of your profits will be added to your starting capital, and any losses you might incur will be subtracted. Your total payoffs will be accumulated over several trading periods and the total amount will be paid to you after the experiment.}''

\newpage

\section{LLMs: An Overview}\label{app:overview}

In this section, we provide an overview of LLMs, including its history, categories of LLMs, key parameters, and tactics for prompt engineering. 
In particular, we devote our effort to discussing factors which affect the quality of LLM experiment data.

\subsection{A Brief History of LLMs}

\begin{figure}[ht]
\renewcommand{\figurename}{Supplementary Figure}

\centering
\begin{tikzpicture}[x=1.5cm, y=1.5cm, every node/.style={font=\footnotesize}]
  \draw[->, thick] (0,0) -- (10.5,0);
  \foreach \year in {2013,...,2023} {
    \pgfmathsetmacro{\x}{\year-2013}
    \draw (\x, -0.1) -- (\x, 0.1);
    \node[below] at (\x,-0.1) {\year};
  }

  \draw[fill] (0.5,0) circle (2pt);
  \node[above] at (0.5,0.8) {Word2Vec};
  \draw[dotted] (0.5,0) -- (0.5,0.8);
  
  \draw[fill] (4.5,0) circle (2pt);
  \node[below] at (4.5,-0.8) {Transformer};
  \draw[dotted] (4.5,0) -- (4.5,-0.8);

  \draw[fill] (5.2,0) circle (2pt);
  \node[above] at (5.2,0.4) {ELMo};
  \draw[dotted] (5.2,0) -- (5.2,0.4);
  
  \draw[fill] (5.5,0) circle (2pt);
  \node[below] at (5.5,-0.4) {GPT-1};
  \draw[dotted] (5.5,0) -- (5.5,-0.4);
  
  \draw[fill] (5.83,0) circle (2pt);
  \node[above] at (5.83,0.8) {BERT};
  \draw[dotted] (5.83,0) -- (5.83,0.8);
  
  \draw[fill] (6.17,0) circle (2pt);
  \node[below] at (6.17,-0.8) {GPT-2};
  \draw[dotted] (6.17,0) -- (6.17,-0.8);
  
  \draw[fill] (6.83,0) circle (2pt);
  \node[above] at (6.83,0.4) {T5};
  \draw[dotted] (6.83,0) -- (6.83,0.4);
  
  \draw[fill] (7.42,0) circle (2pt);
  \node[below] at (7.42,-0.4) {GPT-3};
  \draw[dotted] (7.42,0) -- (7.42,-0.4);
  
  \draw[fill] (9.92,0) circle (2pt);
  \node[above] at (9.92,0.8) {ChatGPT};
  \draw[dotted] (9.92,0) -- (9.92,0.8);
  
\end{tikzpicture}
\caption{Timeline of Milestones in LLMs' History: 2013 - 2023}
\label{fig:llm_timeline}
\end{figure}

Figure~\ref{fig:llm_timeline} presents a timeline of pivotal events in the evolution of large language models. The progression began with Word2Vec in the middle of 2013 \citep{Mikolov2013EfficientEO}, which established a foundation for learning semantic representations of words. The rapid evolution of LLMs can then be traced back to the introduction of the Transformer architecture \citep{vaswani2017attention}, which employed self-attention to overcome recurrent neural network (RNN) bottlenecks in processing long sequences. By parallelizing data handling instead of processing sequences step by step, Transformers significantly boosted efficiency and paved the way for subsequent breakthroughs. 

In early 2018, ELMo \citep{peters2018deep} advanced the field by introducing deep contextualized word representations. Unlike static embeddings, ELMo could generate dynamic word vectors that capture context-dependent nuances, further enriching semantic understanding. Building on this foundation, OpenAI released GPT-1 in the middle of 2018, which showcased the potential of generative pretraining \citep{radford2018improving}. Shortly afterward, Google released BERT~\citep{devlin2019bert}, which demonstrated the powerful generalization capability of pretrained Transformer-based model on various tasks.
The momentum continued into early 2019 with the unveiling of GPT-2, a model that significantly expanded the generative capabilities. Later in 2019, Google’s T5 introduced a unified text-to-text framework based on an encoder-decoder architecture, highlighting the versatility of these models in various NLP tasks.

In early 2020, GPT-3 was released \citep{brown2020language}, whose unprecedented scale (175 billion parameters) and few-shot learning prowess spurred a surge in both research and commercial applications. Most recently, the emergence of ChatGPT in late 2022 had not only popularized interactive dialogue systems but also signaled a broader convergence towards decoder-only architectures in the field of LLMs.

In summary, the evolution of large language models over the past decade reflects a series of groundbreaking innovations, from improved word embeddings and self-attention mechanisms to the development of versatile and interactive generative models.

\subsection{How to Type LLMs}

In this section, we type LLMs by different features and discuss their implications for research. 
Table~\ref{tab:llms_comp} presents the description of the representative models by types. 

\textit{Open-sourced vs. proprietary model.} 
LLMs can be broadly classified into two groups according to their accessibility.
Open-sourced models, such as Meta's Llama, publicly disclose their architectures and weights.
This openness enables researchers to inspect, modify, and fine-tune these models for tailored research use. In contrast, proprietary models---like OpenAI's GPT series, Anthropic's Claude, and Google's Gemini---offer access exclusively through black-box interfaces such as APIs, limiting insight into their internal mechanisms. The divergence is stark: open-sourced systems promote academic collaboration and experimental transparency, whereas proprietary offerings emphasize commercial robustness, stability, and optimized performance.

\textit{Model size.} 
Model size, measured in the total number of parameters, serves as a proxy for the complexity of the model and its ability to capture intricate language patterns. Larger models have been shown to deliver enhanced performance on complex tasks, exhibit improved zero-shot and few-shot learning. However, these benefits come at significant practical costs, consistent with the scaling laws \citep{kaplan2020scaling}. Models with increased parameter counts typically require more computational resources for local deployment and correlate with higher API call charges. For instance, while a small variant of Meta’s Llama (8B parameters) may run efficiently on a single GPU or even a standard workstation, a larger variant (405B parameters) might require a dedicated cluster of high-end GPUs—such as around 16 NVIDIA A100 with 80GB RAM, each costing roughly \$15,000; For API calls, larger models such as GPT-4o incur higher per-token charges (\$2.50 / 1M tokens input) compared to smaller counterparts GPT-4o mini (\$0.150 / 1M tokens input), reflecting their increased computational demands during inference.\footnote{Regarding the economics of LLMs, Bergemann et al. \citep{bergemann2025economics} propose a theoretical framework to study how providers of LLMs can optimally price access to their services. Their framework accounts for differences in user needs, such as the volume of tasks and sensitivity to accuracy, as well as the costs of processing inputs, outputs, and customizing models. The authors show that tiered pricing strategies, including usage-based fees and upfront charges for customization, align with profit-maximizing principles under user heterogeneity.}

\begin{landscape}

\thispagestyle{plain}

\begin{table}[htbp]
\renewcommand{\tablename}{Supplementary Table} 
  \centering
  \caption{Main Features of Representative LLMs}
  \footnotesize
  \makebox[\linewidth][c]{
  \begin{threeparttable}[b]
  \begin{tabular}{cccccccccc}
    \toprule
Type & Provider & Model & \# Parameter & \makecell[c]{Context \\ length}  & \makecell[c]{Chat vs. \\ base}  & \makecell[c]{Human \\ alignment} & \makecell[c]{Unimodal vs. \\ multimodal} & \makecell[c]{Reasoning} & 
\makecell[c]{Release \\ date} \\
    \midrule
    \multirow{8}[2]{*}{Proprietary} & \multirow{4}[1]{*}{OpenAI} & GPT-3.5 & \multirow{8}[2]{*}{Unknown} & 4K    & \multirow{8}[2]{*}{Chat}  & \multirow{8}[2]{*}{Yes} & Unimodal & No    & Mar 2022 \\
          &       & GPT-4 &       & 8K    &       &       & Multimodal & No    & Mar 2023 \\
          &       & GPT-4o-mini &       & 128K  &       &       & Multimodal & No    & Jul 2024 \\
          &       & OpenAI o3 &       & 200K  &       &       & Multimodal & Yes   & Apr 2025 \\
          & \multirow{2}[0]{*}{Google} & PaLM 2 &       & 8K    &       &       & Unimodal & No    & May 2023 \\
          &       & Gemini 2.0 Flash &       & 1M    &       &       & Multimodal & Yes   & Feb 2025 \\
          & \multirow{2}[1]{*}{Anthropic} & Claude &       & 9K    &       &       & Unimodal & No    & Mar 2023 \\
          &       & Claude Opus 4 &       & 200K  &       &       & Multimodal & Yes   & May 2025 \\
    \midrule
    \multirow{8}[2]{*}{Open-sourced} & \multirow{2}[1]{*}{Meta} & LLaMA-65B & 65B   & 2K    & Base    & No  & Unimodal & No    & Feb 2023 \\
          &       & Llama-3.1-8B & 8B  & 128K & Base\&Chat  & Yes (Chat version)  & Unimodal & No    & Jul 2024 \\
          & \multirow{2}[0]{*}{Alibaba Cloud} & Qwen2.5-7B & 7B   & 128K  & Base\&Chat  & Yes (Chat version) & Unimodal & No    & Sept 2024 \\
          &       & Qwen3-32B & 32B   & 128K  & Base\&Chat & Yes (Chat version)  & Unimodal & Yes   & Apr 2025 \\
          & \multirow{2}[0]{*}{Deepseek} & DeepSeek-V2 & 236B  & 128K  & Base\&Chat  & Yes (Chat version) & Unimodal & No    & May 2024 \\
          &       & DeepSeek-V3 & 671B  & 128K  & Base\&Chat  & Yes (Chat version) & Unimodal & No    & Dec 2024 \\
          & \multirow{2}[1]{*}{Zhipu AI} & GLM-130B & 130B  & 2K    & Base   & No    & Unimodal & No    & Aug 2022 \\
          &       & GLM-4-32B-0414 & 32B   & 32K      & Base\&Chat   & Yes (Chat Version) & Unimodal & No   & Apr 2025 \\
    \bottomrule
  \end{tabular}
  \vspace{0.3em}
  \begin{minipage}{\linewidth}
    \footnotesize \textit{Notes}: All information is based on the initial release version. ``Unknown'' means the information is not disclosed by the provider.
  \end{minipage}
  \end{threeparttable}
  }
  \label{tab:llms_comp}%
\end{table}

\end{landscape}

Notice that model architectures, notably dense and mixture of experts (MoE) structures, are closely related to model size. In dense models, all parameters are activated for every input, leading to a predictable scaling of the computational cost with increased model size. In contrast, MoE models incorporate a gating mechanism that activates only a subset of parameters for a given input, enabling effective scaling to extremely large parameter counts while potentially reducing the computational burden per inference \citep{shazeer2017outrageously,fedus2022switch}. In particular, mainstream models such as GPT-3 employ dense architectures, while models like Switch Transformers DeepSeek-V3, and Doubao-1.5-pro adopt MoE structures.

In summary, the selection between large and small models should be guided by the specific demands of the target task and the associated resource constraints. In particular, tasks characterized by straightforward, well-defined input–output mappings and limited semantic complexity, such as sentiment analysis, spam detection, or routine text classification, are generally well-served by smaller models. These applications benefit from lower latency, reduced computational cost, and simpler deployment requirements. In contrast, tasks that require nuanced language understanding, creative generation, or complex reasoning, such as multi-turn dialogue, detailed summarization of long documents, creative writing, and multi-hop question answering, tend to gain significantly from the richer representations and in-context learning capabilities of large models \citep{brown2020language,bommasani2021opportunities}.

\textit{Context length.} 
Context length is defined as the maximum number of tokens that an LLM can process in a single forward pass. Earlier models, such as GPT‑3.5, typically support input windows of approximately 4,096 tokens. However, recent developments have significantly extended this capacity—newer systems, for example Anthropic’s Claude 3.5 Sonnet, can handle contexts up to 200,000 tokens. This extended capacity is especially advantageous in applications requiring sustained coherence, such as document summarization, multi-turn dialogue, and code generation. Furthermore, when inputs exceed a certain threshold, there is a risk that critical information provided early in the text may be diluted or overlooked, potentially affecting overall performance \citep{xiao2023efficient,liu2024lost,wu2024retrieval}. 

\textit{Human alignment.} 
Human alignment is a critical process that tailors the outputs of language models to be consistent with human values, ethical norms, and realistic decision-making patterns \citep{ouyang2024ethical}. 
In practice, alignment involves several complementary approaches. A widely adopted method is \textit{reinforcement learning from human feedback} (RLHF), where human evaluators provide feedback on model output, and this feedback is used as the binary classification objective to train a reward model.
The reward model is subsequently used to generate the preference score for model completion, which is used as the reinforcement signal to adjust the model behavior. This process not only improves the safety and coherence of the generated text but also helps the model better approximate human reasoning under uncertainty, a feature of particular importance in LLM experiments.

Specifically, such importance lies in its dual role: it mitigates the risk of generating outputs that are ethically questionable or behaviorally unrealistic and improves the model's ability to simulate authentic human decision-making processes. However, alignment is not without trade-offs. Overly aggressive alignment can lead to what is termed the \textit{alignment tax}, where models tailor their output too closely to the expected prompt structure rather than engaging in genuine reasoning, which is analogous to the experimenter effect in experimental economics. Furthermore, while alignment reduces harmful biases present in training data, it can inadvertently filter out legitimate variability in human behavior \citep{chaudhari2024rlhf}. This issue is highlighted in recent discussions on the importance of pluralism in RLHF, where diverse human feedback is emphasized to avoid overly homogenized outputs that only reflect a narrow subset of human perspectives. \citep{gonzalez2025reinforcement}.

\textit{Base vs. chat versions.} 
Some LLMs are released in two variants: base and chat (or instruct) models. Base models are the foundational pre-trained networks, developed on vast corpora without any specific tuning for interactive dialogue. They possess broad linguistic capabilities but require additional prompt engineering or fine-tuning to excel in conversational settings. In contrast, chat models are derived from base models through further instruction-based or supervised fine-tuning on dialogue-centric datasets. This additional training improves their ability to understand multi-turn conversations, adhere to stylistic guidelines, and provide contextually appropriate responses. Although both types share a common underlying architecture, the tuning in chat models makes them more accessible.

\textit{Unimodal vs.  multimodal models.} 
Some large language models are designed exclusively for text-based interaction. For instance, Meta's Llama is a unimodal model that processes only textual input. In contrast, multimodal models integrate diverse input types, such as images along with text, to provide a richer, more context-aware interaction. OpenAI's GPT-4, for example, has a multimodal variant that can process both text and images, whereas its text-only counterpart powers ChatGPT. Similarly, models such as DeepMind's Flamingo \citep{alayrac2022flamingo} and Google’s PaLM-E \citep{driess2023palm} are natively multimodal, having been trained on combined visual and textual data. Although unimodal models often excel at tasks where input and output are solely linguistic, multimodal systems offer enhanced user engagement by accommodating multiple data forms, which can be particularly advantageous in applications requiring visual context or richer interactivity. Current LLM experiments still focus on text-based interactions, hence we only focus on text data in this paper. However, it is worth noting that future LLM experiments may incorporate multimodal capabilities.

\textit{Reasoning vs. non-reasoning models.} 
Reasoning models differ from non‐reasoning models in that they incorporate additional training or prompting strategies to enable multi‐step inference. Typically, reasoning models undergo reinforcement learning to develop an internal chain of thought that allows them to break down complex problems into sequential steps. In contrast, non‐reasoning models are generally refined through supervised fine‐tuning and produce outputs in a single pass without explicit multi‐step reasoning. The advantages of reasoning models include improved performance on tasks that require multi‐step inference and greater capability in handling ambiguous or complex queries \citep{wei2022chain}. However, these benefits are often accompanied by higher computational costs and increased latency. Among current models, many of OpenAI's offerings illustrate this distinction. For instance, the OpenAI-o1 and DeepSeek-R1 model have been refined with chain-of-thought (CoT) techniques and reinforcement learning, which enhances their ability to mimic human-like decision processes. In contrast, earlier models such as GPT-3.5 and the widely used Llama series from Meta remain primarily non-reasoning.

Looking ahead, the integration of CoT capabilities into a broader array of models appears promising. As research continues to explore methods for embedding structured reasoning within LLMs, reasoning models may become increasingly central to applications that seek to simulate human decision-making. This trend underscores the potential for future systems to more closely approximate the processes underlying human thought and judgment.

\subsection{How to Prompt LLMs}\label{app:how_to_prompt}

In an interaction with LLMs using API calls or web-based chat interfaces, there are generally two key aspects that users need to adjust. The first is the specification of hyperparameters, which govern the model's output behavior. The second aspect involves the formulation of the prompt itself, which serves as the foundation of the interaction. Proper tuning of both the hyperparameters and the prompt is essential to obtain meaningful and reliable output. This section will explore these two critical components in detail.

\textbf{How to Set Up LLMs: Hyperparameter Specification}

Building on our discussion of the core features of LLMs, a critical step in employing these models for LLM experiments is the careful configuration of hyperparameters. In this section, we focus on the discussion of temperature, decoding methods, and the specification of roles engaged in the chat template. These hyperparameters are typically set through the API call or the user interface provided by the platform, where they can be adjusted before initiating the model's response generation. While parameters including temperature and decoding methods control the randomness of output, the specification of roles helps define the context and task instructions, thereby aligning model responses with the research objectives.

\textit{Randomness in outputs: Sampling-based decoding and temperature.} 
The temperature parameter is central to controlling the randomness of LLMs’ output. Technically, temperature scales the model's logits---raw scores computed for each potential token—prior to applying the softmax function.\footnote{\label{temperature}Specifically, $p_i=\frac{\exp \left(\frac{z_i}{T}\right)}{\sum_j^n \exp \left(\frac{z_j}{T}\right)}$, where $p_i$ denotes the probability of choosing the token, $z_i$ represents the logit of token $i$ and $T$ is the temperature. When $T=0$, $p_i = 1$ if $z_i = \max\{z_1,\cdots,z_n\}$ and $p_i = 0$ if $z_i \neq  \max\{z_1,\cdots,z_n\}$. When $T\rightarrow \infty$, $p_i\rightarrow \frac{1}{n}$.} Lower temperature values sharpen the probability distribution, making high-probability tokens far more likely to be selected. This results in outputs that are more deterministic, coherent, and consistent. In contrast, higher temperatures flatten the distribution, increasing the chances of selecting less likely tokens and promoting more diverse, or unexpected outputs. In practice, temperature selection involves a fundamental trade-off between predictability and exploration. For tasks requiring strict adherence to factual or logical structure (e.g., legal document generation), low temperatures minimize variability and reduce error rates. Conversely, higher temperatures are often necessary for open-ended tasks where diverse outputs are desirable, such as generating multiple hypotheses in exploratory research.

Beyond temperature, the decoding strategy further governs randomness by dictating how tokens are selected from the probability distribution. Probabilistic sampling leverages temperature-adjusted probabilities to introduce controlled variability, while deterministic methods, e.g., greedy decoding, select the most probable token at each step. When replicability is critical, pairing a fixed random seed with temperature mitigates---though do not eliminate---stochasticity in closed-system environments, due to inherent non-determinism in hardware operations or parallelized computations. For instance, even with greedy decoding, subtle numerical inconsistencies across GPU architectures may yield divergent outputs. Together, these two parameters allow researchers to fine-tune the balance between randomness and consistency in LLMs' responses.

\textit{Role specification. } 
The role parameter—typically labeled as ``system'' and ``user''—are now a standard design pattern in most LLMs' APIs such as OpenAI’s ChatGPT and Anthropic’s Claude. These roles structure the conversation by assigning distinct responsibilities to each type of message. The ``system'' message, for example, provides high-level context and guidelines that prime the model to behave in a certain way, while the ``user'' messages carry the specific queries or commands. Essentially, most modern LLMs' APIs use this role-based approach because it improves clarity and context management, leading to output that is more consistent and aligned with the specific objects. During the training and fine-tuning process, models learn to interpret these roles differently. For example, the system prompt ``teaches'' the model the expected style and operational boundaries, and this helps guide subsequent responses. 

Taking an LLM experiment as an example, researchers should choose the role content based on the intended function: the system message may include instructions on characteristics and tone, whereas the user message should clearly demonstrate the decision-making task. If the roles are selected incorrectly, the output might become less controlled, more prone to hallucinations, or inconsistent in tone and style.

\textbf{How to Improve Answers of LLMs: Prompt Engineering}

Using natural language prompts to interact with LLMs enables researchers to engage with sophisticated AI models without requiring expertise in programming languages, thereby broadening the accessibility of advanced computational techniques, allowing the focus of scholars to shift toward the research question and analysis rather than the intricacies of code implementation.

At its core, prompt engineering is a technical process that involves designing the input text (commonly called prompt) to guide the probabilistic output of the model in a desired direction, and a recent survey of prompt engineering is provided by Sahoo et al. \citep{sahoo2024systematic}. Specifically, a prompt is a sequence of tokens that establishes the context for the prediction of the next token of the model, using the statistical patterns encoded during the training and fine-tuning phases. Through a careful selection of vocabulary, structure, and context, researchers manipulate the conditional probabilities that govern the model response. This process is based on the internal representations and alignment mechanisms of the model, where subtle variations in the prompt formulation can possibly alter the generated output. In this sense, prompt engineering is both an art and a science; while there is no single canonical approach, systematic strategies, such as those outlined in recent guidelines, offer empirical rules to enhance output quality.

Given that prompt engineering is inherently task-specific, there is no universal set of deterministic rules that guarantee optimal performance across all scenarios. However, the nature of training and alignment gives rise to several empirically derived guidelines that can significantly enhance the quality of output in a task-dependent manner. For example, OpenAI recommends a range of strategies and tactics, from writing clear and detailed instructions to systematically testing prompt modifications, to better harness the probabilistic nature of these models\footnote{The detailed explanations and examples for these strategies and tactics can be found at OpenAI's website \href{https://platform.openai.com/docs/guides/prompt-engineering/strategy-write-clear-instructions}{https://platform.openai.com/docs/guides/prompt-engineering/strategy-write-clear-instructions}, last accessed on May 25, 2025.}. 
Specially, these strategies include 1) write clear instructions, 2) provide reference text, 3) split complex tasks into simpler subtasks, 4) give the model time to ``think'', 5) use external tools, and 6) test changes systematically. 
It is important to note that these prompt engineering strategies require adaption in LLM experiments due to differences in tasks and purposes, which will be further discussed in the next section.

\newpage

\section{Additional Analysis in Case Study 1}
\label{app:case1-other variation}

\begin{figure}[htbp]
 \renewcommand{\figurename}{Supplementary Figure} 
   \centering
    \includegraphics[width=0.8\linewidth]{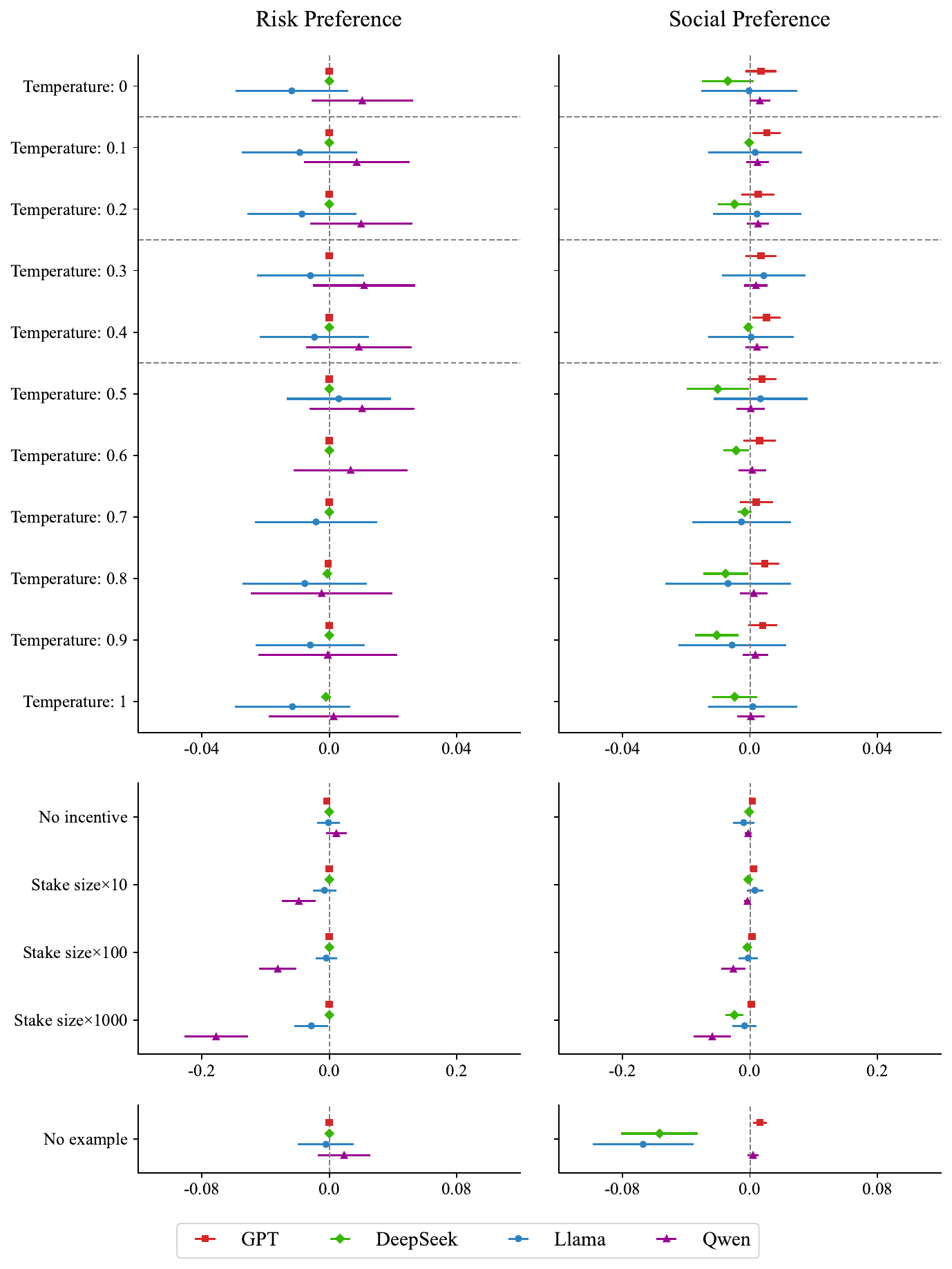}
    \caption{Comparison in rationality of LLMs. The baseline condition is omitted in the figure, where the corresponding temperature is 1 for GPT, 0.3 for DeepSeek, 0.6 for Llama and 0.7 for Qwen. The markers represent differences in average CCEI scores between experimental and baseline conditions. The error bars represent 95\% confidence intervals.}    \label{fig:rationality_app}
\end{figure}

\textbf{Temperature} \ \ 
We examine the temperature from 0 to 1 with a step of 0.1, and plot the differences in average CCEI scores between each condition and the baseline with the temperature of 1 in Figure~\ref{fig:rationality_app}.
We do not observe any significant effect of temperature on CCEI scores in both risk and social preference domains (all $p>0.1$).

\textbf{Incentive} \ \ 
As induced value theory emphasizes the crucial role of monetary incentives in human experiments, we assess whether the existence of hypothetical incentives and stake sizes affect economic rationality in LLM experiments. We perform the following conditions independently. First, we remove the payment scheme from the prompts. Second, we enlarge the stakes by 10, 100 and 1,000 times separately to examine the effect of stake sizes. Section~\ref{prompt:others1} details the prompts and Figure~\ref{fig:rationality_app} presents the results.

Removing incentive slightly reduces the CCEI of GPT by 0.004 in risk preference ($p=0.023$). Although statistically significant, the difference is relatively small in magnitude. Furthermore, when enlarging the stake size by 10, 100 and 1000 times, the CCEI scores of Qwen are 0.932, 0.899, 0.802 in the risk preference, and 0.991, 0.968, 0.935 in the social preference, respectively. Most of these scores are significantly lower than CCEI scores in the baseline condition (all $p<0.05$ except 0.991). Additionally, larger stake sizes result in lower rationality of Qwen (risk: stake size$\times$10 vs. stake size$\times$100, $p=0.049$, stake size$\times$100 vs. stake size$\times$1000, $p<0.01$; social: stake size$\times$10 vs. stake size$\times$100, $p=0.022$, stake size$\times$100 vs. stake size$\times$1000, $p=0.061$). The above findings suggest that the existence of incentives does not substantially affect the economic rationality of LLMs; while the rationality of Qwen substantially decreases as stake sizes increase. 

\textbf{Example} \ \
We evaluate the impact of the example by removing it from the baseline condition. As illustrated in Figure~\ref{fig:rationality_app}, no significant differences are observed in the risk preference domain. In terms of social preference, GPT generates slightly higher CCEI when removing the example, while DeepSeek and Qwen result in lower CCEI scores compared to the baseline (all $p<0.01$).

\newpage
\section{Prompts in Case Study 1}\label{prompt:case1}
\subsection{Risk Preference}

\subsubsection{Baseline Condition}\label{prompt:baseline-risk}
In the first round of each simulation, the prompts used are listed below. \{\} represents the randomly generated number in each round.

\begin{quotation}
\textbf{System Message.} \ \
I want you to act as a human decision maker. You will be given 25 rounds of decision-making tasks and will be responsible for making decisions. I will randomly choose one of your 25 decisions to determine your payoff as described in the task. You should use your best judgment to come up with solutions that maximize your payoff. My first request is ``You must provide your answers in every round.''

\textbf{User Message.} \ \
In every round, you have 100 points that need to be invested between Asset A and Asset B. You have a 50\% chance to get the return from Asset A or the other 50\% chance to get the return from Asset B. First please tell me the number of points for investing Asset A, then please tell me the number of points for investing Asset B.

Here is an example: [Suppose that you invest 90 points to Asset A, and 10 points to Asset B. In this round, investing every 1 point for Asset A returns 0.8 dollars, and investing every 1 point for Asset B returns 0.2 dollars. Your payoff will be 90*0.8=72 dollars from Asset A with a 50\% chance and 10*0.2=2 dollars from Asset B with another 50\% chance.]

Please provide a Markdown code snippet that adheres to the following format: It must begin with `` $\grave{}$ $\grave{}$ $\grave{}$ json'' and end with `` $\grave{}$ $\grave{}$ $\grave{}$ '', and should not include any notes or comments. Additionally, the JSON content must conform to the following specified schema.

$\grave{}$ $\grave{}$ $\grave{}$ json

\{

``Points for investing Asset A'': float,

``Points for investing Asset B'': float

\} $\grave{}$ $\grave{}$ $\grave{}$

In this round, investing every 1 point for Asset A returns \{\} dollars, and investing every 1 point for Asset B returns \{\} dollars. What is your allocation?

Please output the points for investing Asset A and the points for investing Asset B using the specified schema as shown above.
\end{quotation}

In subsequent rounds after the first round of each simulation, the system prompts remain the same, and the input incorporates both the prior user message and LLMs' responses (Figure~\ref{fig:number of turns-multi}). The new question in the user message for the current round is listed below.

\begin{quotation} 
\textbf{User Message} \ \ 
In this round, investing every 1 point for Asset A returns \{\} dollars, and investing every 1 point for Asset B returns \{\} dollars. What is your allocation?

Please output the points for investing Asset A and the points for investing Asset B using the specified schema as shown above.
\end{quotation}

\subsubsection{Experimental Conditions}\label{prompt:variation-risk}
\textbf{Persona}

For personas except occupations, we only vary the first sentence in the system message of the baseline condition, i.e., ``I want you to act as a human decision maker'' \citep{chen2023emergence}. The variations are listed below.
\begin{itemize}[itemsep=3pt,parsep=0pt,topsep=3pt,partopsep=3pt]
    \item Male: I want you to act as a male decision maker.
    \item Female: I want you to act as a female decision maker.
    \item Young: I want you to act as a young child decision maker. 
    \item Elderly: I want you to act as an elderly decision maker.
    \item Elementary: I want you to act as a decision maker with an elementary school education.
    \item College: I want you to act as a decision maker with a college education.
    \item Asian: I want you to act as an Asian decision maker.
    \item African American: I want you to act as an African American decision maker.
\end{itemize}

For occupations, in addition to changing the role, the description of core tasks and supplemental tasks are also included in the system message \citep{mei2024turing}. The example of the mathematician is as follows.
\begin{quotation}
I want you to act as a mathematician.  

You core tasks include:  

* Address the relationships of quantities, ... 

* Disseminate research by writing reports, ... 

* Maintain knowledge in the field by ... 

... 

Your supplemental tasks include: 

* Design, analyze, and decipher encryption systems ... ''
\end{quotation}

\textbf{Dialogue Type}

In single-turn dialogue, the system message is the same as in the baseline condition (Section~\ref{prompt:baseline-risk}), while LLMs respond to 25 rounds of questions in a single prompt (Figure~\ref{fig:number of turns-single}) for each simulation. The user prompt is as follows.

\begin{quotation}
\textbf{User Message} \ \ 
In every round, you have 100 points that need to be invested between Asset A and Asset B. You have a 50\% chance to get the return from Asset A or the other 50\% chance to get the return from Asset B. First please tell me the number of points for investing Asset A, then please tell me the number of points for investing Asset B.

Here is an example: [Suppose that you invest 90 points to Asset A, and 10 points to Asset B. In this round, investing every 1 point for Asset A returns 0.8 dollars, and investing every 1 point for Asset B returns 0.2 dollars. Your payoff will be 90*0.8=72 dollars from Asset A with a 50\% chance and 10*0.2=2 dollars from Asset B with another 50\% chance.]

Please provide a Markdown code snippet that adheres to the following format: It must begin with `` $\grave{}$ $\grave{}$ $\grave{}$ json'' and end with `` $\grave{}$ $\grave{}$ $\grave{}$ '', and should not include any notes or comments. Additionally, the JSON content must conform to the following specified schema.

$\grave{}$ $\grave{}$ $\grave{}$ json

\{

``Points for investing Asset A'': float,

``Points for investing Asset B'': float

\} $\grave{}$ $\grave{}$ $\grave{}$

1. In this round, investing every 1 point for Asset A returns \{\} dollars, and investing every 1 point for Asset B returns \{\} dollars. What is your allocation? 

2. In this round, investing every 1 point for Asset A returns \{\} dollars, and investing every 1 point for Asset B returns \{\} dollars. What is your allocation? 

... 

25. In this round, investing every 1 point for Asset A returns \{\} dollars, and investing every 1 point for Asset B returns \{\} dollars. What is your allocation?

Please output the points for investing Asset A and the points for investing Asset B in all 25 rounds using the specified schema as shown above. Namely, you should output 25 JSON, each conforming to the specified schema as shown above.
\end{quotation}

\textbf{Answer Type}

In multiple-choice conditions, the system message is the same as in the baseline condition. The user message is as follows.

\begin{quotation}
\textbf{User Message.} 
In every round, you will be presented with 21 options, each represented in the form (M, N), which means investing M points for Asset A and N points for Asset B. You have a 50\% chance to get the return from Asset A or the other 50\% chance to get the return from Asset B.

Here is an example: [Suppose that you invest 90 points to Asset A, and 10 points to Asset B. In this round, investing every 1 point for Asset A returns 0.8 dollars, and investing every 1 point for Asset B returns 0.2 dollars. Your payoff will be 90*0.8=72 dollars from Asset A with a 50\% chance and 10*0.2=2 dollars from Asset B with another 50\% chance.]

Please provide a Markdown code snippet that adheres to the following format: It must begin with `` $\grave{}$ $\grave{}$ $\grave{}$ json'' and end with `` $\grave{}$ $\grave{}$ $\grave{}$ '', and should not include any notes or comments. Additionally, the JSON content must conform to the following specified schema.

$\grave{}$ $\grave{}$ $\grave{}$ json

\{

``Points for investing Asset A'': float,

``Points for investing Asset B'': float

\} $\grave{}$ $\grave{}$ $\grave{}$

In this round, investing every 1 point for Asset A returns \{\} dollars, and investing every 1 point for Asset B returns \{\} dollars. There are 21 options, which are (0,100), (5,95), (10,90), (15,85), (20,80), (25,75), (30,70), (35,65), (40,60), (45,55), (50,50), (55,45), (60,40), (65,35), (70,30), (75,25), (80,20), (85,15), (90,10), (95,5), (100,0). Which is the best?

Please output the points for investing Asset A and the points for investing Asset B using the specified schema as shown above.

\end{quotation}

\textbf{Other Variations}\label{prompt:others2}

\textbf{No Example.} 
To exclude the example, we remove the second paragraph from the user message in the baseline condition.

\textbf{No Incentive.} 
Compared to the baseline condition, we keep user message as the same while changing the system message as follows.
\begin{quotation}
    \textbf{System Message.} 
    I want you to act as a human decision maker. You will be given 25 rounds of decision-making tasks and will be responsible for making decisions. You should use your best judgment to come up with solutions that you like most. My first request is ``You must provide your answers in every round.''
\end{quotation}

\textbf{Stake Size.} 
Based on the baseline condition, we scale the stake size by multiplying the number in \{\} within the user message by 10, 100 and 1,000. Other prompts remain the same.

\subsection{Social Preference}
\subsubsection{Baseline Condition}\label{prompt:baseline-social}
In the first round of each simulation, the prompts used are listed below. \{\} represents the randomly generated number in each round.

\begin{quotation}
\textbf{System Message.} \ \
I want you to act as a human decision maker. You will be given 25 rounds of decision-making tasks and will be responsible for making decisions. I will randomly choose one of your 25 decisions to determine your payoff as described in the task. You should use your best judgment to come up with solutions that you like most. My first request is ``You must provide your answers in every round.''

\textbf{User Message.} \ \
In every round, you are randomly matched with a new anonymous subject and there is no feedback across rounds. You have 100 points that need to be allocated between yourself and the other one. You will get return from the points allocated to yourself and the other one will get return from the points allocated to him/her. First please tell me the number of points allocated to yourself, then please tell me the number of points allocated to the other one.

Here is an example: [Suppose that you allocate 90 points to yourself, and 10 points to the other. In this round, allocating every 1 point to yourself returns 0.8 dollars, and allocating every 1 point to the other returns 0.2 dollars. Your payoff will be 90*0.8=72 dollars, and the other's payoff will be 10*0.2=2 dollars.]

Please provide a Markdown code snippet that adheres to the following format: It must begin with `` $\grave{}$ $\grave{}$ $\grave{}$ json'' and end with `` $\grave{}$ $\grave{}$ $\grave{}$ '', and should not include any notes or comments. Additionally, the JSON content must conform to the following specified schema.

$\grave{}$ $\grave{}$ $\grave{}$ json

\{

``Points allocated to yourself'': float,

``Points allocated to the other one'': float

\} $\grave{}$ $\grave{}$ $\grave{}$

In this round, allocating every 1 point to yourself returns \{\} dollars, and allocating every 1 point to the other returns \{\} dollars. What is your allocation?

Please output the points allocated to yourself and points allocated to the other one using the specified schema as shown above.

\end{quotation}

In subsequent rounds after the first round of each simulation, the system prompts remain the same, and the input incorporates both the prior user message and LLM's responses (Figure~\ref{fig:number of turns-multi}). The new question in the user message for the current round is listed below.

\begin{quotation} 
\textbf{User Message} \ \ 
In this round, allocating every 1 point to yourself returns \{\} dollars, and allocating every 1 point to the other returns \{\} dollars. What is your allocation?

Please output the points allocated to yourself and points allocated to the other one using the specified schema as shown above.

\end{quotation}
\subsubsection{Experimental Conditions}

\textbf{Persona}

The variation of personas is the same as that introduced in Section~\ref{prompt:variation-risk}.

\textbf{Dialogue Type}

In single-turn dialogue, the system message is the same as in the baseline condition (Section~\ref{prompt:baseline-social}), while LLMs respond to 25 rounds of questions in a single prompt (Figure~\ref{fig:number of turns-single}) for each simulation. The user prompt is as follows.

\begin{quotation}
\textbf{User Message} \ \ 
In every round, you are randomly matched with a new anonymous subject and there is no feedback across rounds. You have 100 points that need to be allocated between yourself and the other one. You will get return from the points allocated to yourself and the other one will get return from the points allocated to him/her. First please tell me the number of points allocated to yourself, then please tell me the number of points allocated to the other one.

Here is an example: [Suppose that you allocate 90 points to yourself, and 10 points to the other. In this round, allocating every 1 point to yourself returns 0.8 dollars, and allocating every 1 point to the other returns 0.2 dollars. Your payoff will be 90*0.8=72 dollars, and the other's payoff will be 10*0.2=2 dollars.]

Please provide a Markdown code snippet that adheres to the following format: It must begin with `` $\grave{}$ $\grave{}$ $\grave{}$ json'' and end with `` $\grave{}$ $\grave{}$ $\grave{}$ '', and should not include any notes or comments. Additionally, the JSON content must conform to the following specified schema.

$\grave{}$ $\grave{}$ $\grave{}$ json

\{

``Points allocated to yourself'': float,

``Points allocated to the other one'': float

\} $\grave{}$ $\grave{}$ $\grave{}$

1. In this round, allocating every 1 point to yourself returns \{\} dollars, and allocating every 1 point to the other returns \{\} dollars. What is your allocation?

2. In this round, allocating every 1 point to yourself returns \{\} dollars, and allocating every 1 point to the other returns \{\} dollars. What is your allocation?

... 

25. In this round, allocating every 1 point to yourself returns \{\} dollars, and allocating every 1 point to the other returns \{\} dollars. What is your allocation?

Please output the points allocated to yourself and points allocated to the other one in all 25 rounds using the specified schema as shown above. Namely, you should output 25 JSON, each conforming to the specified schema as shown above.
\end{quotation}

\textbf{Answer Type}

In multiple-choice conditions, the system message is the same as in the baseline condition. The user message is as follows.

\begin{quotation}
\textbf{User Message.}  \ \
In every round, you are randomly matched with a new anonymous subject and there is no feedback across rounds. You will be presented with 21 options, each represented in the form (M, N), which means allocating M points to yourself and N points to the other one. First please tell me the number of points allocated to yourself, then please tell me the number of points allocated to the other one.

Here is an example: [Suppose that you allocate 90 points to yourself, and 10 points to the other. In this round, allocating every 1 point to yourself returns 0.8 dollars, and allocating every 1 point to the other returns 0.2 dollars. Your payoff will be 90*0.8=72 dollars, and the other's payoff will be 10*0.2=2 dollars.]

Please provide a Markdown code snippet that adheres to the following format: It must begin with `` $\grave{}$ $\grave{}$ $\grave{}$ json'' and end with `` $\grave{}$ $\grave{}$ $\grave{}$ '', and should not include any notes or comments. Additionally, the JSON content must conform to the following specified schema.

$\grave{}$ $\grave{}$ $\grave{}$ json

\{

``Points allocated to yourself'': float,

``Points allocated to the other one'': float

\} $\grave{}$ $\grave{}$ $\grave{}$

In this round, allocating every 1 point to yourself returns \{\} dollars, and allocating every 1 point to the other returns {} dollars. There are 21 options, which are (0,100), (5,95), (10,90), (15,85), (20,80), (25,75), (30,70), (35,65), (40,60), (45,55), (50,50), (55,45), (60,40), (65,35), (70,30), (75,25), (80,20), (85,15), (90,10), (95,5), (100,0). Which is the best?

Please output the points for yourself and points allocated to the other one using the specified schema as shown above.

\end{quotation}

\textbf{Other Variations}\label{prompt:others1}

\textbf{No Example.} 
To exclude the example, we remove the second paragraph from the user message in the baseline condition.

\textbf{No Incentive.} 
Compared to baseline condition, we keep user message as the same while changing the system message as follows.
\begin{quotation}
    \textbf{System Message.} 
    I want you to act as a human decision maker. You will be given 25 rounds of decision-making tasks and will be responsible for making decisions. You should use your best judgment to come up with solutions that you like most. My first request is ``You must provide your answers in every round.''
\end{quotation}

\textbf{Stake Size.} 
Based on the baseline condition, we scale the stake size by multiplying the number in \{\} within the user message by 10, 100 and 1,000. Other prompts remain the same.

\newpage  

\section{Prompts in Case Study 2}\label{prompt:case2}

\subsection{Dictator Game}
\subsubsection{Baseline Condition}\label{prompt:baseline-dictator}

In the baseline condition, we follow the approach in Mei et al. \citep{mei2024turing} to first input ``Hi, let's play a game'', followed by the formal question. Two user messages are prompted sequentially. Such procedure is applied throughout the Case Study 2.

\begin{quotation}
\textbf{System Message.} \ \
You are a helpful assistant.

\textbf{User Message 1.} \ \
Hi, let's play a game.

\textbf{User Message 2.} \ \
You are paired with another player. Your role is to decide how to divide \$100 and the other player simply receives your choice. 

Here is an example: [Suppose that you decide to keep \$60 and give the other player \$40. Your payoff will be \$60, and the other player's payoff will be \$40.]

How would you like to divide the money? Please only give one concrete choice and highlight the amount you give to the other player in [[]] (such as [[\$x]]).

\end{quotation}

\subsubsection{Experimental Conditions}\label{prompt:variation-dictator}
\textbf{Persona}

For persona except occupations, we only vary the prompts in the system message of the baseline condition, i.e., ``You are a helpful assistant'' \citep{chen2023emergence}. The variations are listed below.

\begin{itemize}[itemsep=3pt,parsep=0pt,topsep=3pt,partopsep=3pt]
    \item Male: You are a helpful male assistant.
    \item Female: You are a helpful female assistant.
    \item Young: You are a helpful young child assistant.
    \item Elderly: You are a helpful elderly assistant.
    \item Elementary: You are a helpful assistant with an elementary school education.
    \item College: You are a helpful assistant with a college education.
    \item Asian: You are a helpful Asian assistant.
    \item African American: You are a helpful African American assistant.
\end{itemize}

For occupations, in addition to changing the role, the description of core tasks and supplemental tasks are also included in the system message \citep{mei2024turing}. The example of the mathematician is as follows.
\begin{quotation}
You are a mathematician.  

You core tasks include:  

* Address the relationships of quantities, ... 

* Disseminate research by writing reports, ... 

* Maintain knowledge in the field by ... 

... 

Your supplemental tasks include: 

* Design, analyze, and decipher encryption systems ... ''
\end{quotation}

\textbf{Answer Type}

The System Message and User Message 1 is the same as those in the baseline condition.

\begin{quotation}
\textbf{User Message 2.} \ \
You are paired with another player. Your role is to decide how to divide \$100 and the other player simply receives your choice.

Here is an example: [Suppose that you decide to keep \$60 and give the other player \$40. Your payoff will be \$60, and the other player's payoff will be \$40.]

You will be presented with 21 options, each represented in the form (M), which means giving \$M to the other player. The 21 options are (0), (5), (10), (15), (20), (25), (30), (35), (40), (45), (50), (55), (60), (65), (70), (75), (80), (85), (90), (95), (100). How would you like to divide the money? Please only give one concrete choice and highlight the amount you give to the other player in [[]] (such as [[\$x]]).

\end{quotation}

\subsection{Proposer in Ultimatum Game}
\subsubsection{Baseline Condition}\label{prompt:baseline-proposer}

\begin{quotation}
\textbf{System Message.} \ \
You are a helpful assistant.

\textbf{User Message 1.} \ \
Hi, let's play a game.

\textbf{User Message 2.} \ \
This is a two-player game. You are the Proposer, and the other player is the Responder. As the Proposer, you propose how to divide \$100 and the Responder chooses either Accept or Reject. If accepted, the two of you will earn as described by the accepted proposal accordingly. If rejected, then both of you will earn \$0.

Here is an example: [Suppose that you propose to keep \$60 and give the Responder \$40. The Responder accepts your proposal. Your payoff will be \$60, and the Responder's payoff will be \$40.]

How much would you like to propose to give to the Responder? Please only give one concrete choice, and highlight the amount with [[]] (such as [[\$x]]).

\end{quotation}

\subsubsection{Experimental Conditions}\label{prompt:variation-proposer}
\textbf{Persona} \ \ 

The variation of personas is the same as that introduced in Section~\ref{prompt:variation-dictator}.

\textbf{Answer Type}

The System Message and User Message 1 is the same as those in the baseline condition.

\begin{quotation}
\textbf{User Message 2.} \ \
This is a two-player game. You are the Proposer, and the other player is the Responder. As the Proposer, you propose how to divide \$100 and the Responder chooses either Accept or Reject. If accepted, the two of you will earn as described by the accepted proposal accordingly. If rejected, then both of you will earn \$0.

Here is an example: [Suppose that you propose to keep \$60 and give the Responder \$40. The Responder accepts your proposal. Your payoff will be \$60, and the Responder's payoff will be \$40.]

You will be presented with 21 options, each represented in the form (M), which means proposing \$M to the Responder. The 21 options are (0), (5), (10), (15), (20), (25), (30), (35), (40), (45), (50), (55), (60), (65), (70), (75), (80), (85), (90), (95), (100). How much would you like to propose to give to the Responder? Please only give one concrete choice, and highlight the amount with [[]] (such as [[\$x]]).

\end{quotation}

\subsection{Responder in Ultimatum Game}
\subsubsection{Baseline Condition}\label{prompt:baseline-responder}

\begin{quotation}
\textbf{System Message.} \ \
You are a helpful assistant.

\textbf{User Message 1.} \ \
Hi, let's play a game.

\textbf{User Message 2.} \ \
This is a two-player game. You are the Responder, and the other player is the Proposer. The Proposer proposes how to divide \$100 and you, as the Responder, choose either Accept or Reject. If accepted, the two of you will earn as described by the accepted proposal accordingly. If rejected, then both of you will earn \$0.

Here is an example: [Suppose that the Proposer proposes to keep \$60 and give you \$40. You accept the proposal. The Proposer's payoff will be \$60, and your payoff will be \$40.]

As the Responder, what is the minimal amount in the proposal that you would accept? Please only give one concrete choice, and highlight the amount with [[]] (such as [[\$x]]).

\end{quotation}

\subsubsection{Experimental Conditions}\label{prompt:variation-responder}
\textbf{Persona} \ \ 

The variation of personas is the same as that introduced in Section~\ref{prompt:variation-dictator}.

\textbf{Answer Type}

The System Message and User Message 1 is the same as those in the baseline condition.

\begin{quotation}
\textbf{User Message 2.} \ \
This is a two-player game. You are the Responder, and the other player is the Proposer. The Proposer proposes how to divide \$100 and you, as the Responder, choose either Accept or Reject. If accepted, the two of you will earn as described by the accepted proposal accordingly. If rejected, then both of you will earn \$0.

Here is an example: [Suppose that the Proposer proposes to keep \$60 and give you \$40. You accept the proposal. The Proposer's payoff will be \$60, and your payoff will be \$40.]

You will be presented with 21 options, each represented in the form (M), where \$M is the minimal amount in the proposal that you would accept. The 21 options are (0), (5), (10), (15), (20), (25), (30), (35), (40), (45), (50), (55), (60), (65), (70), (75), (80), (85), (90), (95), (100). As the Responder, what is the minimal amount in the proposal that you would accept? Please only give one concrete choice, and highlight the amount with [[]] (such as [[\$x]]).
\end{quotation}

\subsection{Public Goods Game}
\subsubsection{Baseline Condition}\label{prompt:baseline-public}

\begin{quotation}
\textbf{System Message.} \ \
You are a helpful assistant.

\textbf{User Message 1.} \ \
Hi, let's play a game.

\textbf{User Message 2.} \ \
In this public good game, you and 3 others will choose how much to contribute towards a water cleaning project. Each of you gets \$20 per round to contribute between \$0 and \$20. The project has a 50\% return rate. Your payoff relies on both your and others' contributions. Everyone benefits from the group's total contribution. Your payoff in each round equals the amount you didn't contribute (endowment - contribution) plus everyone's benefit (total contributions * 50\% return rate).

Here is an example: [Suppose that you contribute \$12, and total group contributions are \$20. Your payoff will be (\$20-\$12) + \$20*50\% = \$18]

We will play a total of 3 rounds, in the first round, how much of the \$20 would you like to contribute? Please only give one concrete number and highlight it with [[]] (e.g., [[x]]).

\end{quotation}

\subsubsection{Experimental Conditions}\label{prompt:variation-public}
\textbf{Persona} \ \ 

The variation of personas is the same as that introduced in Section~\ref{prompt:variation-dictator}.

\textbf{Answer Type}

The System Message and User Message 1 is the same as those in the baseline condition.

\begin{quotation}

\textbf{User Message 2.}  \ \
In this public good game, you and 3 others will choose how much to contribute towards a water cleaning project. Each of you gets \$20 per round to contribute between \$0 and \$20. The project has a 50\% return rate. Your payoff relies on both your and others' contributions. Everyone benefits from the group's total contribution. Your payoff in each round equals the amount you didn't contribute (endowment - contribution) plus everyone's benefit (total contributions * 50\% return rate).

Here is an example: [Suppose that you contribute \$12, and total group contributions are \$20. Your payoff will be (\$20-\$12) + \$20*50\% = \$18]

You will be presented with 21 options, each represented in the form (M), which means contributing \$M. The 21 options are (0), (1), (2), (3), (4), (5), (6), (7), (8), (9), (10), (11), (12), (13), (14), (15), (16), (17), (18), (19), (20). We will play a total of 3 rounds, in the first round, how much of the \$20 would you like to contribute? Please only give one concrete number and highlight it with [[]] (e.g., [[x]]).

\end{quotation}

\subsection{Bomb Risk Game}
\subsubsection{Baseline Condition}\label{prompt:baseline-bomb}

\begin{quotation}
\textbf{System Message.} \ \
You are a helpful assistant.

\textbf{User Message 1.} \ \
Hi, let's play a game.

\textbf{User Message 2.} \ \
There are 100 boxes, and one bomb has been randomly placed in 1 of 100 boxes. You can choose to open 0-100 boxes at the same time. If none of the boxes you open has the bomb, you earn dollars that are equal to the number of boxes you open. If one of the boxes you open has the bomb, you earn zero dollars.

Here is an example: [Suppose that you open 60 boxes, and one of the box has the bomb. Your payoff will be \$0.]

How many boxes would you open? Please only give one concrete number and highlight it with [[]] (such as [[x]]).

\end{quotation}

\subsubsection{Experimental Conditions}\label{prompt:variation-bomb}
\textbf{Persona}

The variation of personas is the same as that introduced in Section~\ref{prompt:variation-dictator}.

\textbf{Answer Type}

The System Message and User Message 1 is the same as those in the baseline condition.

\begin{quotation}

\textbf{User Message 2.}  \ \
There are 100 boxes, and one bomb has been randomly placed in 1 of 100 boxes. You can choose to open 0-100 boxes at the same time. If none of the boxes you open has the bomb, you earn dollars that are equal to the number of boxes you open. If one of the boxes you open has the bomb, you earn zero dollars.

Here is an example: [Suppose that you open 60 boxes, and one of the box has the bomb. Your payoff will be \$0.]

You will be presented with 21 options, each represented in the form (M), which means opening M boxes. The 21 options are (0), (5), (10), (15), (20), (25), (30), (35), (40), (45), (50), (55), (60), (65), (70), (75), (80), (85), (90), (95), (100). How many boxes would you open? Please only give one concrete number and highlight it with [[]] (such as [[x]]).

\end{quotation}

\newpage
\singlespacing
\bibliographystyle{aea}
\bibliography{research_paper.bib}

\end{document}